\documentclass[aps,twocolum,amsmath,amssymb,a4paper,allowtoday
,accepted=2025-07-14
]{quantumarticle}
\pdfoutput=1

\usepackage{amsmath,braket,amssymb}
\usepackage{amsthm}
\usepackage[final]{graphicx}
\usepackage{subfigure}
\usepackage{xcolor}
\usepackage[english]{babel}
\usepackage{color}
\usepackage{mathtools}
\usepackage{empheq}
\usepackage{tensor}

\usepackage{hhline}

\usepackage[colorlinks,citecolor=darkBlue,linkcolor=darkBlue,
urlcolor=blue,hyperindex]{hyperref}

\usepackage[normalem]{ulem}
\normalfont

\usepackage[linesnumbered, ruled,vlined]{algorithm2e}

\graphicspath{{./images/},{./imagesAppendix/}}

\definecolor{forestgreen}{rgb}{0.08, 0.4, 0.13}
\definecolor{darkBlue}{rgb}{0.08, 0.13, 0.4}

\definecolor{THc}{rgb}{0.9,0.3,0.2}

\newcommand\numberthis{\addtocounter{equation}{1}\tag{\theequation}}

\newcommand{\revA}[1]{{#1}}

\newcommand{\idg}[1]{{\bfseries #1)}}

\renewcommand{\eqref}[1]{Eq.~(\ref{#1})} 

\newcommand{\subfigimg}[3][,]{%
	\setbox1=\hbox{\includegraphics[#1]{#3}}%
	\leavevmode\rlap{\usebox1}%
	\rlap{\hspace*{2pt}\raisebox{\dimexpr\ht1-0.5\baselineskip}{{\bfseries \large\textsf{#2}}}}%
	\phantom{\usebox1}%
}

\begin{document}

\title{Probing quantum complexity via universal saturation of stabilizer entropies}

\author{Tobias Haug}
\email{tobias.haug@u.nus.edu}
\affiliation{Quantum Research Center, Technology Innovation Institute, Abu Dhabi, UAE}
\affiliation{Blackett Laboratory, Imperial College London SW7 2AZ, UK}
\author{Leandro Aolita}
\affiliation{Quantum Research Center, Technology Innovation Institute, Abu Dhabi, UAE}
\author{M.S. Kim}
\affiliation{Blackett Laboratory, Imperial College London SW7 2AZ, UK}

\date{\today}
	
\begin{abstract}
Nonstabilizerness or `magic' is a key resource for quantum computing and a necessary condition for quantum advantage.
Non-Clifford operations turn stabilizer states into resourceful states,  where the amount of nonstabilizerness is quantified by resource measures such as stabilizer R\'enyi entropies (SREs). 
Here, we show that SREs saturate their maximum value at a critical number of non-Clifford operations. Close to the critical point SREs show universal behavior. Remarkably,  the derivative of the SRE crosses at the same point independent of the number of qubits and can be rescaled onto a single curve.
We find that the critical point depends non-trivially on R\'enyi index $\alpha$.
For random Clifford circuits doped with T-gates, the critical T-gate density scales independently of $\alpha$. In contrast, for random Hamiltonian evolution, the critical time scales linearly with qubit number for $\alpha>1$, while is a constant for $\alpha<1$. 
This highlights that $\alpha$-SREs reveal fundamentally different aspects of nonstabilizerness depending on $\alpha$: $\alpha$-SREs with $\alpha<1$ relate to Clifford simulation complexity, while $\alpha>1$ probe the distance to the closest stabilizer state and approximate state certification cost via Pauli measurements. 
As technical contributions, we observe that the Pauli spectrum of random evolution can be approximated by two highly concentrated peaks which allows us to compute its SRE. Further, we introduce a class of random evolution that can be expressed as random Clifford circuits and rotations, where we provide its exact SRE.
Our results opens up new approaches to characterize the complexity of quantum systems.
\end{abstract}
	
\maketitle

\section{Introduction}

Nonstabilizerness or `magic' has been proposed as the resource theory of fault-tolerant quantum computers~\cite{veitch2014resource}. It lower bounds the non-Clifford resources needed to run quantum computers~\cite{howard2017application} and  relates to the complexity of classical simulation algorithms based on Clifford operations~\cite{bravyi2016trading}.

Nonstabilizerness monotones are non-increasing under Clifford operations, while applying non-Clifford operations can enhance the amount of nonstabilizerness of a state. On a quantitative level, the relationship between quantum complexity and number of non-Clifford operations is an intensely studied topic~\cite{haferkamp2022efficient,liu2022many,leone2021quantum}.
As nonstabilizerness is a measure of complexity for quantum computers and a necessary condition for quantum advantage, it is paramount to understand how nonstabilizerness depends on the number of non-Clifford operations~\cite{white2021conformal,ellison2021symmetry,sarkar2020characterization,liu2022many}.

A wide range of magic monotones have been proposed, such as stabilizer rank~\cite{bravyi2016trading}, min-relative entropy of magic~\cite{bravyi2019simulation,liu2022many} and log-free robustness of magic~\cite{howard2017application,liu2022many}.  These measures probe different aspects of nonstabilizerness. In particular, the log-free robustness of magic can be related to the complexity of a classical simulation algorithm, while the min-relative entropy of magic is related to the distance to the closest stabilizer state. However, they require an optimization program to be computed, making them in general intractable for the study of larger system sizes. Thus, methods to feasibly approximate these monotones are desired.

$\alpha$-Stabilizer R\'enyi entropies (SREs) have been proposed to quantitatively explore nonstabilizerness harnessing their efficient numerical~\cite{leone2022stabilizer,haug2022quantifying,lami2023quantum}, analytic~\cite{lopez2024exact,turkeshi2023pauli} and experimental accessibility~\cite{haug2022scalable,oliviero2022measuring,haug2023efficient,bluvstein2024logical,haug2025efficientwitnessingtestingmagic}. Here, the R\'enyi index $\alpha$ indicates the moment of the SRE. They are monotones for pure states for $\alpha\geq2$~\cite{leone2024stabilizer}, while monotonicity can be violated for $\alpha<2$~\cite{haug2023stabilizer}. SREs relate to phases of error-corrected circuits~\cite{niroula2023phase}, quantify the entanglement spectrum~\cite{tirrito2023quantifying}, bound fidelity estimation~\cite{leone2023phase,leone2023nonstabilizerness,hinsche2024efficient}, characterize the robustness of shadow tomography~\cite{brieger2023stability} and characterize pseudorandom states~\cite{gu2023little,haug2023pseudorandom}. Further, SREs characterize many-body phenomena~\cite{gu2024doped} such as phase transitions~\cite{haug2022quantifying,tarabunga2023many}, frustration~\cite{odavic2022complexity}, 
random matrix product states~\cite{frau2024non,lami2024quantum,paviglianiti2024estimating,mello2024hybrid},
localization~\cite{turkeshi2023measuring} and out-of-time-order correlators~\cite{leone2021quantum,garcia2022resource,leone2023nonstabilizerness}. SREs also lower bound other intractable magic monotones~\cite{haug2023stabilizer,haug2023efficient}.

Here, we study $\alpha$-SREs for random Clifford circuits doped with T-gates, as well as random Hamiltonian evolution in time.
We reveal that SREs saturate their maximal value at a critical T-gate density $q_{\text{c},\alpha}$ or critical time $t_{\text{c},\alpha}$ in the thermodynamic limit. Around the critical point, the the SRE become universal, which we demonstrate by the derivative of the SRE crossing at the critical point for all number of qubits $N$. Further, by a simple rescaling we can collapse the derivative to a single curve, hinting at a possible connection to phase transitions. 

For random Clifford circuits doped with T-gates, SREs grow linearly with number of T-gates until reaching a critical T-gate density which is independent of number of qubits $N$ for any $\alpha$.
In contrast, for random Hamiltonian evolution $\alpha<1$ SREs increase exponentially at short times, becoming extensive already at $1/\text{poly}(N)$ time. In contrast, $\alpha>1$ SREs grow independent of $N$ with a critical time that is linear with $N$, which we find to be a tight lower bound of the stabilizer fidelity.

We argue that $\alpha>1$ and $\alpha<1$ probe different aspects of nonstabilizerness and relate to different quantum computational tasks.
$\alpha<1$ SREs (especially $\alpha=1/2$) relate to the number of superpositions of stabilizer states needed to represent a given state which characterizes the cost of Clifford-based simulation algorithms as well as cost of fault-tolerant state preparation. In contrast, $\alpha>1$ SREs can be seen as the distance to the closest stabilizer state which for example characterizes the cost of Pauli-based fidelity certification. 

We argue that Clifford simulation and state certification become inefficient at the same T-gate doping for Clifford circuits. In contrast, for random evolution these two tasks have completely distinct timescales. In particular, Clifford simulation of random Hamiltonian evolution becomes difficult beyond exponentially small times, while (approximate) state certification is possible up to constant times. 

Thus, while both Clifford simulation and state certification with Pauli measurements are intrinsically linked to nonstabilizerness, the efficiency of these tasks is in general not correlated.  Their efficiency depends on two very different aspects of nonstabilizerness, which can be probed with $\alpha$-SREs. 

As technical contributions, we show that random Hamiltonian evolution has a Pauli spectrum with two distinct peaks, which allows us to compute its SRE. We also introduce a type of random evolution which can be expressed as Clifford circuits with small rotations, which possesses an analytic form of the SRE for $\alpha=2$.

\begin{figure*}[htbp]
	\centering	
	\subfigimg[width=0.95\textwidth]{}{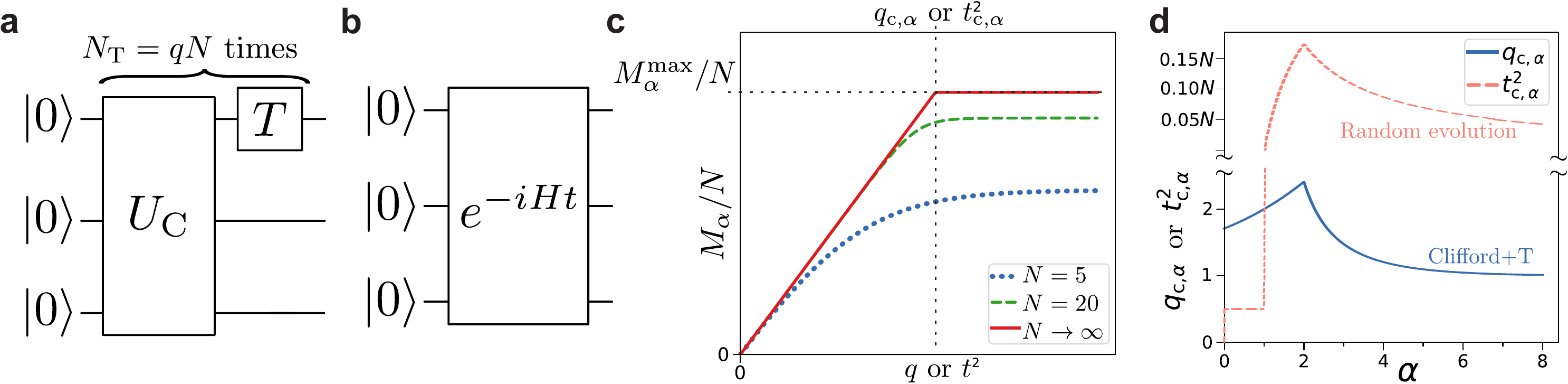}
	\caption{Overview of main results. We study two models: \idg{a} Circuits composed of randomly chosen Clifford circuits $U_\text{C}$ doped with $N_\text{T}=qN$ T-gates, where $N$ is the number of qubits and $q$ the T-gate density. 
 \idg{b} Evolution in time $t$ for random Hamiltonians  starting from an initial stabilizer state.
 \idg{c} $\alpha$-stabilizer R\'enyi entropy (SRE) $M_\alpha$ increases monotonously with $q$ and $t$ until converging to a constant. For large $N$ we observe  a sharp transition to maximal SRE $M_\alpha^\text{max}$ (horizontal dashed line) for a critical T-gate density $q_{\text{c},\alpha}$ or critical time $t_{\text{c},\alpha}$ (vertical dashed line). The derivative of $M_\alpha$ crosses at the critical point for all $N$ and the dynamics close to the critical point can be mapped onto a single curve (see Fig.\ref{fig:CliffordT_renyi} and Fig.~\ref{fig:randomEvol_renyi2}).
 \idg{d} Critical T-gate density $q_{\text{c},\alpha}$ and time $t_{\text{c},\alpha}$ as function of R\'enyi index $\alpha$. Both $q_{\text{c},\alpha}$ and $t_{\text{c},\alpha}$ vary non-monotonously with $\alpha$ as they probe different aspects of nonstabilizerness complexity. In particular, critical time $t_{\text{c},\alpha}$ changes its scaling from constant ($\alpha<1$) to  $t_{\text{c},\alpha}\sim \sqrt{N}$ ($\alpha>1$). 
	}
	\label{fig:overview}
\end{figure*}

Our main results are summarized in Fig.~\ref{fig:overview}, while the critical T-gate density and time are shown in Table~\ref{tab:saturation}.

\begin{table}[htbp]\centering
\begin{tabular}{|c|c| }
                \hline
     Index & Random Clifford+T \\
                     \hline
                     $\alpha=0$ & $q_{\text{c},0}=1$ \\
$\alpha\le2$ & $q_{\text{c},\alpha}\approx (1-\alpha)\frac{\ln(2)}{\ln(2^{-\alpha}+\frac{1}{2})}$\\
        $\alpha=2$ & $q_{\text{c},2}=\log(2)/\log(\frac{4}{3})$ \\
    $\alpha>2$ & $q_{\text{c},\alpha}\approx -\frac{\ln(2)}{\ln(2^{-\alpha}+\frac{1}{2})}$\\
                    \hline
     & Random evolution \\
                    \hline
$\alpha=0$ & $t_{\text{c},\alpha}^2=0$\\
    $0<\alpha<1$ & $t_{\text{c},\alpha}^2\approx\frac{1}{2}$\\
    $1<\alpha\le2$ & $t_{\text{c},\alpha}^2\approx-\frac{1-\alpha}{2\alpha}N\ln(2)$\\
    $\alpha>2$ & $t_{\text{c},\alpha}^2\approx\frac{1}{2\alpha}N\ln(2)$\\ 
    \hline
\end{tabular}
\caption{Critical T-gate density $q_{\text{c},\alpha}$ for random Clifford circuits doped with T-gates and critical time $t_{\text{c},\alpha}$ for evolution with random Hamiltonians. At $q_{\text{c},\alpha}$ and $t_{\text{c},\alpha}$, SREs converge to their maximal value for $N\gg1$. }
\label{tab:saturation}
\end{table}

\section{Stabilizer R\'enyi entropy}\label{sec:SE}
For an $N$-qubit state $\ket{\psi}$, the $\alpha$-SRE is given by~\cite{leone2022stabilizer}
\begin{equation}\label{eq:SRE}
M_{\alpha}(|\psi\rangle)=(1-\alpha)^{-1} \ln(2^{-N} \sum_{\sigma \in \mathcal{P} }\braket{\psi|\sigma|\psi}^{2\alpha})\,.
\end{equation}
where $\alpha$ is the index of the SRE and $\mathcal{P}=\{\sigma\}_{\sigma}$ is the set of all $4^N$ Pauli strings $\sigma\in\{I,X,Y,Z\}^{N}$ which are tensor products of Pauli matrices. 
We note the limit $\alpha\rightarrow1$, which can be easily shown using l'H\^opital's rule and is called the von Neumann stabilizer entropy~\cite{haug2023stabilizer}
\begin{equation}\label{eq:SEneumann}
    M_1(\ket{\psi})=-2^{-N}\sum_{\sigma\in \mathcal{P}}\braket{\psi|\sigma|\psi}^{2}\ln(\braket{\psi|\sigma|\psi}^{2})\,.
\end{equation}
For convenience, we also define the SRE density 
\begin{equation}
    m_\alpha=M_\alpha/N\,.
\end{equation}
$M_\alpha$ is a resource measure of nonstabilizerness for pure states~\cite{veitch2014resource}. SREs are faithful, i.e. $M_\alpha=0$ only for pure stabilizer states, while $M_\alpha>0$ for all other states. Further, $M_\alpha$ is a monotone for $\alpha\geq2$~\cite{leone2024stabilizer}, i.e. non-increasing under free operations, which are Clifford operations that map pure states to pure states. Note that for $\alpha<2$ SREs can violate the monotonicity condition~\cite{haug2023stabilizer}. For all $\alpha$, SREs are invariant under Clifford unitaries $U_\text{C}$, i.e. $M_\alpha(U_\text{C} \ket{\psi})=M_\alpha(\ket{\psi})$.
SREs are also additive with $M_\alpha(\ket{\psi}\otimes\ket{\phi})=M_\alpha(\ket{\psi})+M_\alpha(\ket{\phi})$. 
Note that $M_\alpha$ are not strong monotones for any $\alpha$~\cite{haug2023stabilizer}. 
As R\'enyi entropies, SREs are monotonously increasing with decreasing $\alpha$, i.e.
\begin{equation}
N\ln(2)\ge M_\alpha\ge M_{\alpha'}\ge0
\end{equation}
for $\alpha<\alpha'$.

As other nonstabilizerness monotones, we also consider the min-relative entropy of magic~\cite{bravyi2019simulation,liu2022many}
\begin{equation}\label{eq:dmin}
	D_\text{min}(\ket{\psi})=-\log\left(\max_{\ket{\phi}\in\text{STAB}}\vert\braket{\psi\vert\phi}\vert^2\right)\,,
\end{equation}
where the maximum is taken over the set of pure stabilizer states. Here, $F_\text{STAB}=\max_{\ket{\phi}\in\text{STAB}}\vert\braket{\psi\vert\phi}\vert^2$ is the stabilizer fidelity. $D_\text{min}$ measures the distance between $\ket{\psi}$ and its closest stabilizer state. It is upper bounded by $D_\text{min}\le N\ln(2)$, which is asymptotically reached for Haar random states~\cite{liu2022many}. 
The SRE lower bounds $D_\text{min}$ for $N>1$
\begin{equation}\label{eq:boundDmin}
D_{\rm min}(\ket{\psi}) \geq \frac{\alpha-1}{2\alpha}M_\alpha(\ket{\psi})  \, \qquad (\alpha> 1) \,.
\end{equation}
We also consider the log-free robustness of magic~\cite{howard2017application,liu2022many}
\begin{equation}
	\text{LR}(\rho)=\log\left[\text{min}_x\left\{ \sum_i \vert x_i \vert: \rho=\sum_i x_i \eta_i\right\}\right]\,,
\end{equation}
where $S=\{\eta_i \}_i$ is the set of pure $N$-qubit stabilizer states. 
The SRE lower bounds the log-free robustness for $\alpha\ge1/2$~\cite{howard2017application,leone2022stabilizer}
\begin{equation}\label{eq:mn_lr}
	{\rm LR}(\ket{\psi}) \geq \frac{1}{2}M_{\alpha}(\ket{\psi}) \, \qquad (\alpha\geq 1/2)\,.
\end{equation}
Finally, the SRE is a lower bound to the stabilizer nullity $\nu\ge M_\alpha$ which is given by
\begin{equation}
    \nu(\ket{\psi})=N\ln(2)-\ln(s(\ket{\psi}))\,
\end{equation}
where $s(\ket{\psi})=\vert\{\sigma : \bra{\psi}\sigma\ket{\psi}^2=1\}\vert$ being the size of the set of all Pauli operators that stabilize $\ket{\psi}$.

\section{Clifford circuits with T-gates}\label{sec:CliffT}
We now study the SRE for random circuits composed of Clifford unitaries and T-gates~\cite{haferkamp2022efficient}.
We consider a circuit of $N_\text{T}$ layers consisting of randomly sampled Clifford circuits $U_\text{C}^{(k)}$ and the single-qubit T-gate $T=\text{diag}(1,\exp(-i\pi/4)$
\begin{equation}\label{eq:random_CliffT}
\ket{\psi(N_\text{T})}=U_\text{C}^{(0)}[\prod_{k=1}^{N_\text{T}} (T\otimes I_{N-1}) U_\text{C}^{(k)} ]\ket{0}\,.
\end{equation}
For $N_\text{T}=0$, we have Clifford states, while for $N_\text{T}\sim N$ we have highly random states~\cite{leone2021quantum}.

\begin{figure*}[htbp]
	\centering	
	\subfigimg[width=0.26\textwidth]{a}{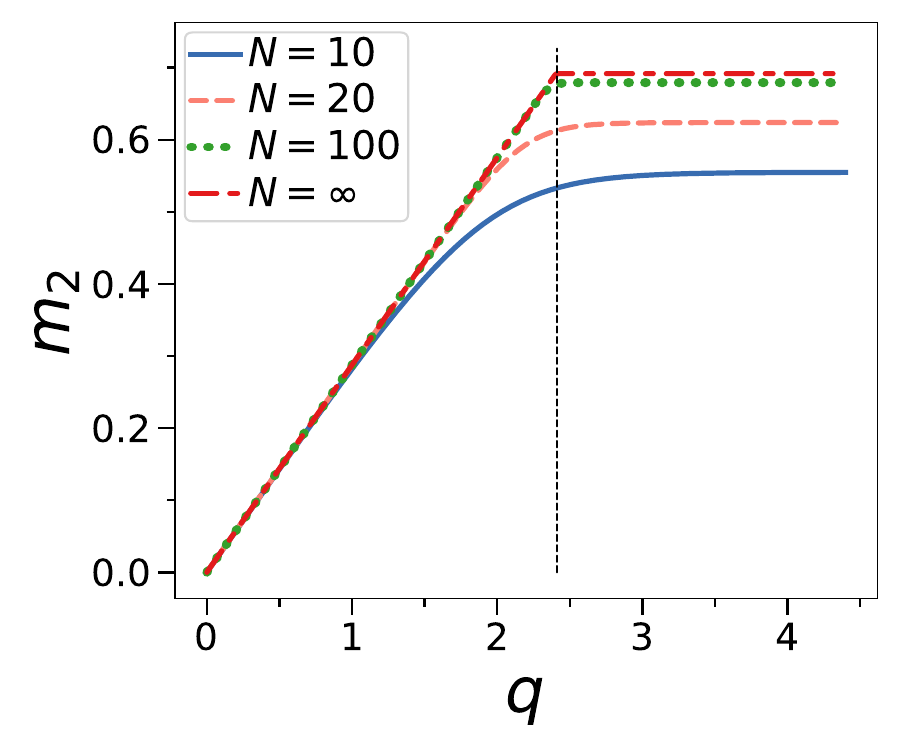}
 	\subfigimg[width=0.26\textwidth]{b}{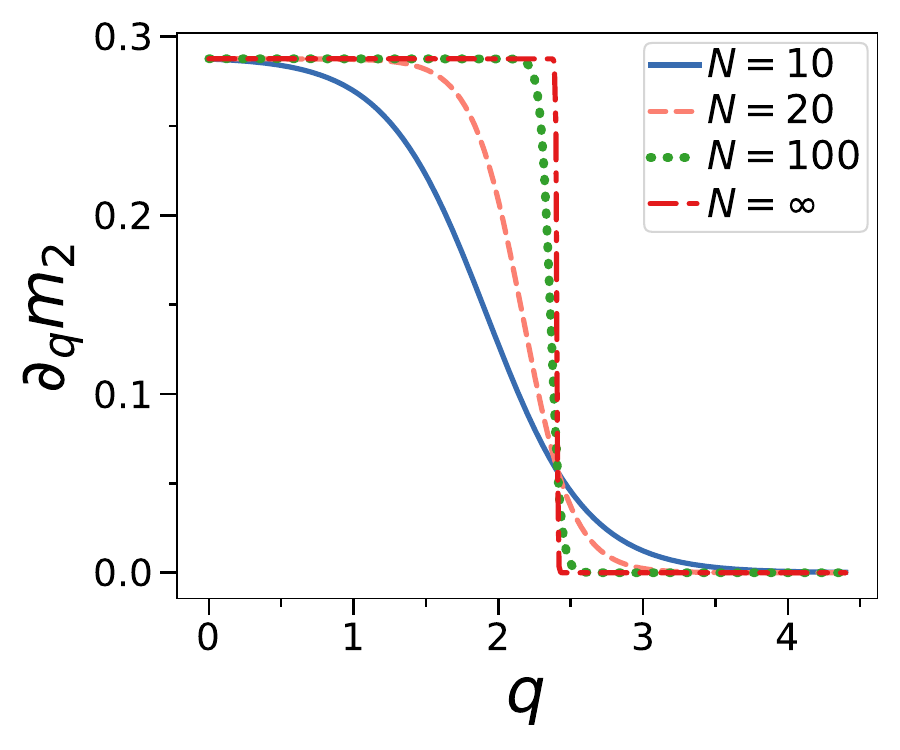}
 	\subfigimg[width=0.26\textwidth]{c}{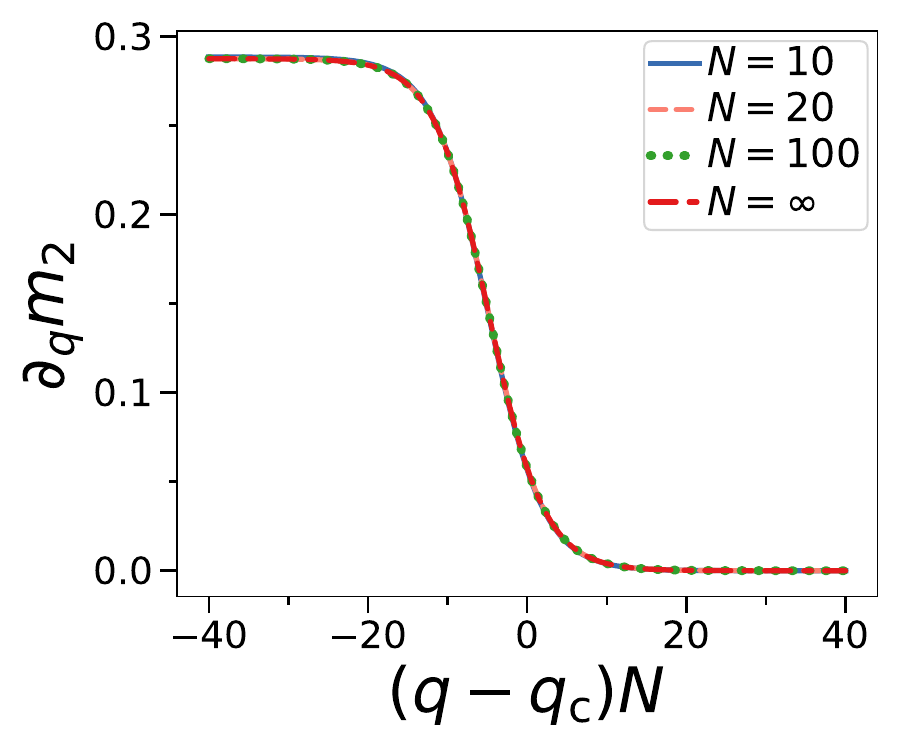}
	\caption{Universal behavior of $2$-SRE for random Clifford + T circuits. \idg{a} $m_2$ against T-gate density $q=N_\text{T}/N$ for different $N$. Black vertical dashed line is transition T-gate density $q_{\text{c},2}=\ln(2)/\ln(4/3)$ \idg{b} Derivative of SRE in respect to T-gate density $\partial_q m_2$ against $q$ with universal crossing for all $N$ at critical $q_{\text{c},2}=\ln(2)/\ln(\frac{4}{3})$.
 \idg{c} Collapse of $\partial_q m_2$ against $q$ when shifted by $q_{\text{c},2}$ and scaled by $N$. 
 Close to the critical density $q_{\text{c},2}$, curves for different $N$ intersect at a single point and can be mapped onto each other by a simple rescaling, which is hallmark of universality.
	}
	\label{fig:CliffordT_renyi}
\end{figure*}

\emph{Analytic SRE.}
The average SRE of such states is known exactly for $\alpha=2$~\cite{leone2022stabilizer}
\begin{equation}\label{eq:exact_renyi}
M_2(N_\text{T})=-\ln\left[\frac{4+(2^N-1)\left(\frac{-4+3(4^N-2^N)}{4(4^N-1)}\right)^{N_\text{T}}}{(3+2^N)}\right]
\end{equation}
For $N\gg1$, this simplifies to
\begin{equation}\label{eq:m2approx}
M_2(q)\approx-\ln[4\times 2^{-N}+\left(\frac{3}{4}\right)^{qN}]\,,
\end{equation}
where we defined the $T$-gate density $q=N_\text{T}/N$.
We study the SRE density $m_2=M_2/N$ in Fig.\ref{fig:CliffordT_renyi}a for different $N$. We observe that $m_2$ increases linearly with $q$ and converges to a constant for large $q$. For large $N$, we observe that the convergence appears to be a sharp transition to the maximal SRE $m_2=\ln(2)$ which occurs at a critical T-gate density $q_{\text{c},2}$. We determine $q_{\text{c},2}$ by studying the scaling at finite $N$~\cite{osterloh2002scaling}. We find that $\partial_q m_2$ as function of $q$ exhibits scale-invariant properties, i.e. the curves for different $N$ can be mapped onto each other by appropriate rescaling around the critical point, a hallmark for phase transitions~\cite{osterloh2002scaling}. In particular, we find that $\partial_q m_2$ intersects for all $N$ at the same point, which gives us $q_{\text{c},2}$. Using~\eqref{eq:m2approx}, we find that the intersection occurs for all $N$ at the critical T-gate density
\begin{equation}
q_{\text{c},2}=\ln(2)/\ln(\frac{4}{3})\approx2.40942\,.
\end{equation} 
In Fig.\ref{fig:CliffordT_renyi}b we plot the derivative $\partial_q m_2$, observing that  for all $N$ the curves indeed intersect at $q_{\text{c},2}$. 
In Fig.\ref{fig:CliffordT_renyi}c, we observe that by shifting $q$ with $q_{\text{c},2}$ and rescaling with $N$, we can collapse the curves of different $N$, as expected for the scale-invariant behavior close to critical points~\cite{osterloh2002scaling}. 

Next, we investigate the case $\alpha=0$
\begin{equation}\label{eq:M0}
M_0(\ket{\psi})=\ln\Big(\sum_\sigma\Theta(\bra{\psi}\sigma\ket{\psi}^2)\Big)-N\ln(2)
\end{equation}
where $\Theta(x)$ is the Heaviside function with $\Theta(x)=0$ for $x\le0$ and $\Theta(x)=1$ for $x>0$. 
Stabilizer states are stabilized by a commuting subgroup $G$ of $2^N$ Pauli strings with $\bra{\psi}\sigma \ket{\psi}^2=1$ for $\sigma\in G$. The group $G$ has $N$ generators.  Applying a T-gate on a stabilizer state breaks at most one generator of $G$, resulting in a state with $N-1$ generators and $2^{N-1}$ stabilizing Pauli strings. In fact, each additional T-gate can break only one generator. Thus, we find $M_0\le N_\text{T} \ln(2)$ and $M_0\le N\ln(2)$~\cite{leone2022stabilizer}.  When the T-gate is applied after a random Clifford circuit, the $T$-gate will break one of the generators of $G$ with overwhelming probability. Thus, with overwhelming probability $N_\text{T}\approx N$ T-gates are necessary and sufficient to reach $M_0=0$, thus the critical T-gate density is $q_{\text{c},0}=1$.

\emph{Approximation of SRE.}
We now provide an estimate for the transition for other $\alpha$. A single T-gate applied on a Clifford state gives an SRE of $M_\alpha^T=(1-\alpha)^{-1}\ln(2^{-\alpha}+\frac{1}{2})$.  
For $\alpha=0$, each additional T-gate increases $M_0$ by the same amount $M_0^T$, yielding $M_0= N_\text{T}M_0^T$ until reaching the maximum $M_0^\text{max}= N_\text{T}M_0^T$ for $N_\text{T}=N$.

\begin{figure}[b]
	\centering	
	\subfigimg[width=0.235\textwidth]{a}{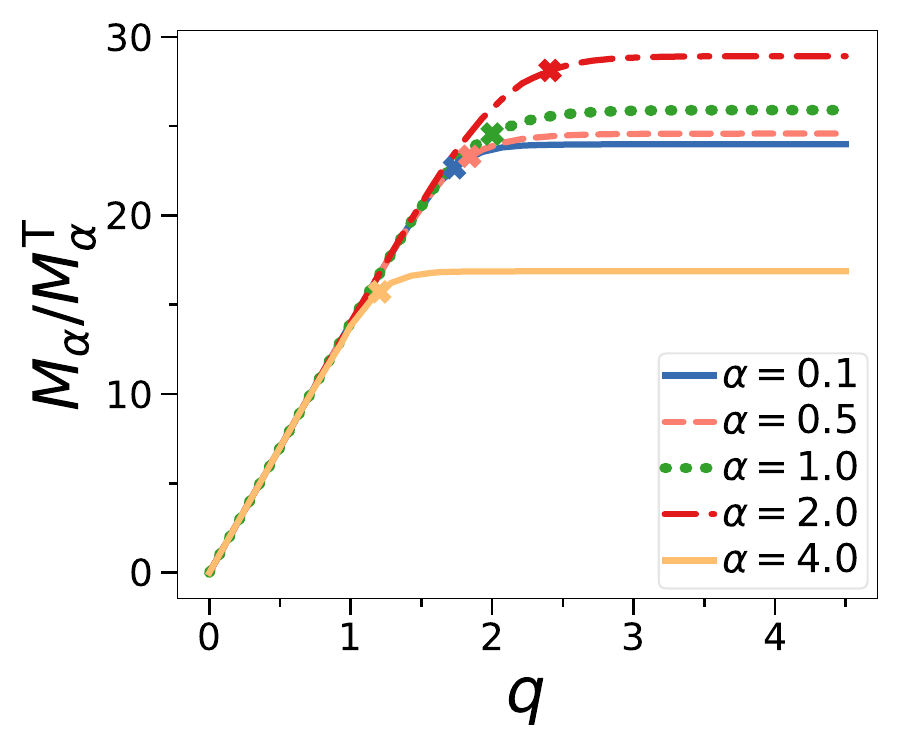}
 	\subfigimg[width=0.235\textwidth]{b}{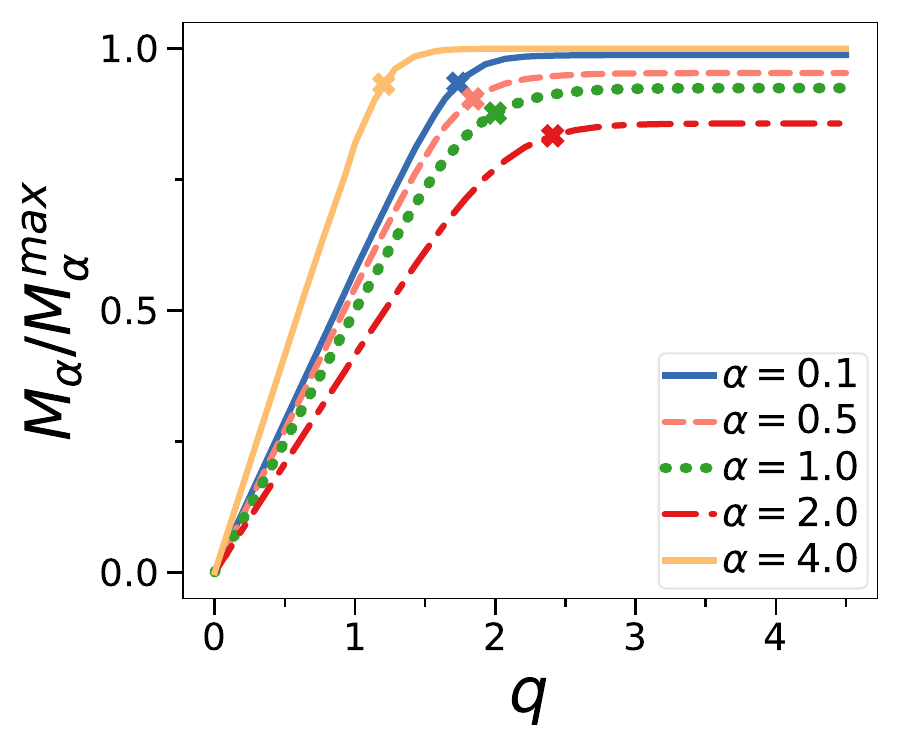}
	\caption{ Saturation of $\alpha$-SREs for random Clifford circuits doped with T-gates. 
 \idg{a} $M_\alpha$ divided by $\alpha$-SRE per T-gate $M_\alpha^\text{T}$ against T-gate density $q$. The crosses denote the critical T-gate density $q_{\text{c},\alpha}$ as derived in~\eqref{eq:transitionTgate}.
  \idg{b} $M_\alpha$ divided by maximal SRE $M_\alpha^\text{max}$ from~\eqref{eq:renyimax} derived for $N\gg1$. We show $N=14$ qubits \revA{and average over 20 random instances, where we note that due to self-averaging the variance in SRE is small.}
	}
	\label{fig:SRET}
\end{figure}

We find a similar analytic relationship for $\alpha=2$. In particular, for large $N\gg1$,~\eqref{eq:exact_renyi} shows that each additional T-gate increases $M_2$ by $M_2^T$  until the SRE is maximal. Thus, we have $M_2\approx N_\text{T}M_2^T$, where $N_\text{T}$ is the number of applied T-gates.  As shown in Fig.~\ref{fig:SRET}a, we observe numerically that this linear relationship between $M_\alpha$ and $N_\text{T}$  also applies for other $\alpha$, i.e.
\begin{equation}\label{eq:SRECliffordNT}
    M_\alpha(N_\text{T})\approx N_\text{T}M_\alpha^T\,.
\end{equation}

The critical T-gate density $q_{\text{c},\alpha}$ is reached when the SRE becomes maximal, which we can approximate with
\begin{equation}\label{eq:getcrit}
    q_{\text{c},\alpha}\approx\frac{1}{N} M_{\alpha}^\text{max}/M_\alpha^T\,.
\end{equation}
Next, we estimate the value of the maximal SRE for $N\gg1$. First, we recall the fact that nonstabilizerness approaches the asymptotically the maximal possible value for randomly chosen states~\cite{liu2022many}. 
Thus, by estimating the SRE of a random state for large $N$ we can approximate $M_{\alpha}^\text{max}$. A random state is spread out over the whole Pauli spectrum where for simplicity we assume a uniform distribution with $\bra{\psi}\sigma\ket{\psi}^2=2^{-N}$ for $\sigma \in \mathcal{P}/\{I\}$.  While such a state is not a positive density matrix, we find from numerical studies that this spectrum is a sufficiently good approximation of an actual random Pauli spectrum for large $N$. \revA{In fact, we find the concentration around our predicted value for highly random states as shown in Fig.~\ref{fig:histPauli}a for $t=1.5$. } 
Using our ansatz, we find
\begin{align*}
    M_{\alpha}^\text{max}&\approx M^\text{uniform}_\alpha=(1-\alpha)^{-1}\ln(2^{-N}\sum_{\sigma\in\mathcal{P}}\bra{\psi}\sigma\ket{\psi}^{2\alpha})\\
    &= (1-\alpha)^{-1}\ln(2^{-N}(1+(4^N-1)2^{-N\alpha})\\
    &\approx (1-\alpha)^{-1}\ln(2^{-N}+2^{N(1-\alpha)})
\end{align*}
\revA{As further confirmation, our result matches $M_\alpha^\text{Haar}$ of Haar random states up to constant corrections~\cite{turkeshi2023pauli}, where Haar random states are known to have asymptotically maximal nonstabilizerness~\cite{liu2022many} (see Appendix~\ref{sec:compHaar} for explicit comparison).} 
Now, in the limit $N\gg1$ we find
\begin{align*}
\alpha\le2: \,\,\,M_{\alpha}^\text{max}\approx& N\ln(2)\\
\alpha>2: \,\,\,M_{\alpha}^\text{max}\approx& -(1-\alpha)^{-1}N\ln(2)\numberthis\label{eq:renyimax}
\end{align*}
Note that our result matches numerical simulations shown in Fig.~\ref{fig:SRET}b and the analytical values known for $\alpha=0$ and $\alpha=2$. 
We can now compute the critical T-gate density by inserting~\eqref{eq:renyimax} into~\eqref{eq:getcrit}
\begin{align*}
\alpha\le2: \,\,\,q_{\text{c},\alpha}&\approx (1-\alpha)\frac{\ln(2)}{\ln(2^{-\alpha}+\frac{1}{2})}\\
    \alpha>2: \,\,\,q_{\text{c},\alpha}&\approx -\frac{\ln(2)}{\ln(2^{-\alpha}+\frac{1}{2})}\,.\numberthis\label{eq:transitionTgate}
\end{align*}

\section{Random basis evolution}\label{sec:Pauli}
 
We proceed to investigate another circuit model which can be seen as a type of random time evolution. It consist of a deep Clifford circuit as in \eqref{eq:random_CliffT} with many layers $d$, but replace the T-gates with a parameterized rotation $R_z(\theta)=\exp(-i\theta/2 \sigma_z)$. A similar model with randomly chosen $\theta$ has been shown to produce highly random states~\cite{haah2024efficient}.
Here, instead we choose very small $\theta$. In particular, we choose $\theta=2t/\sqrt{d}$ with $d\gg N$
\begin{equation}\label{eq:random_CliffGUE}
\ket{\psi_\text{c}(t)}=U_\text{C}^{(0)}\prod_{k=1}^{d} \left(R_z\left(\frac{2t}{\sqrt{d}}\right)\otimes I_{N-1}\right) U_\text{C}^{(k)} \ket{0}\,.
\end{equation}
Here, we can interpret $t$ as evolution time.
We argue that this circuit model describes a type of (time-dependent) random evolution: In particular, if we regard one layer, it consists of transformation into a random basis  with Clifford $U_\text{C}^{(k)}$ and $z$-rotation in the transformed basis by small angle $\theta=2t/\sqrt{d}$. 
This can be seen as a kind of a trotterized evolution with time-dependent Hamiltonian $H(t)$ which rapidly changes between different bases.
We find numerically that the dynamics matches  closely the evolution time of random Hamiltonians as shown in Appendix~\ref{sec:GUEApp}. 

\begin{figure*}[htbp]
	\centering	
	\subfigimg[width=0.26\textwidth]{a}{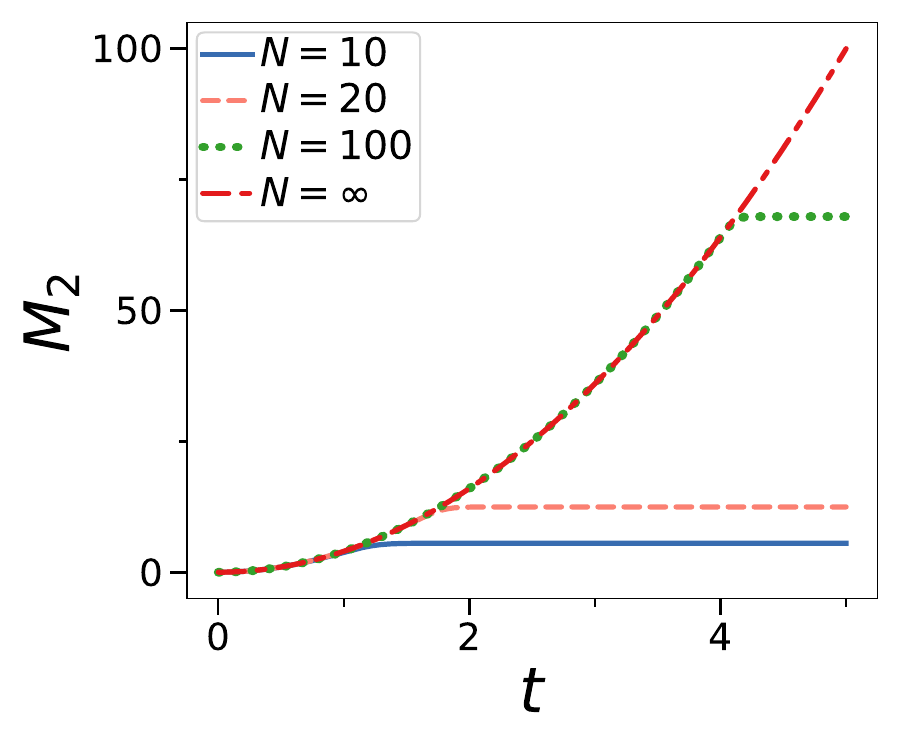}
 	\subfigimg[width=0.26\textwidth]{b}{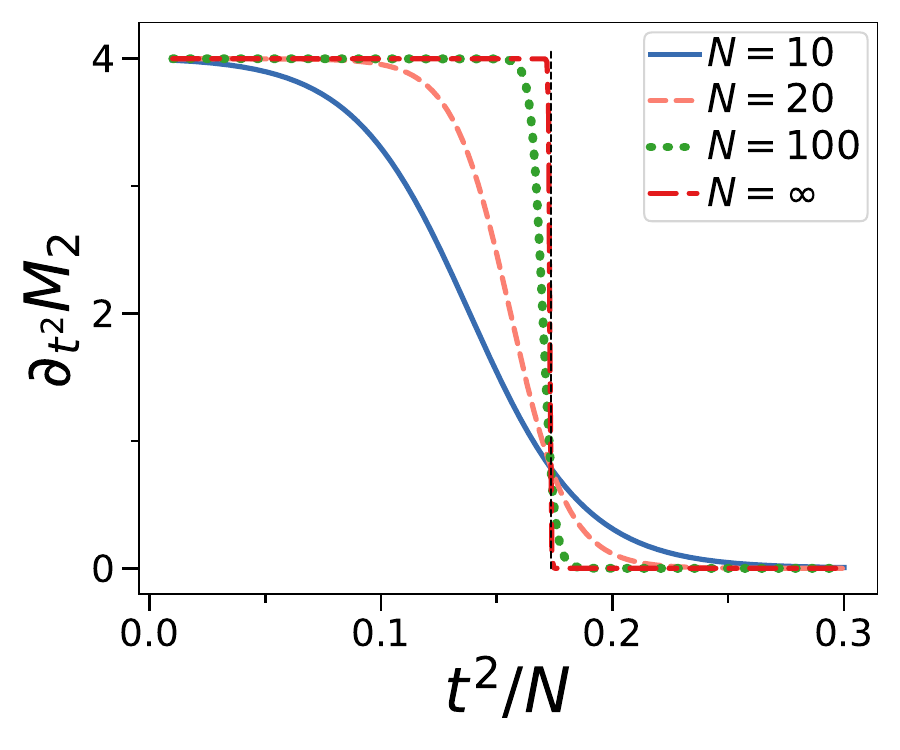}
 	\subfigimg[width=0.26\textwidth]{c}{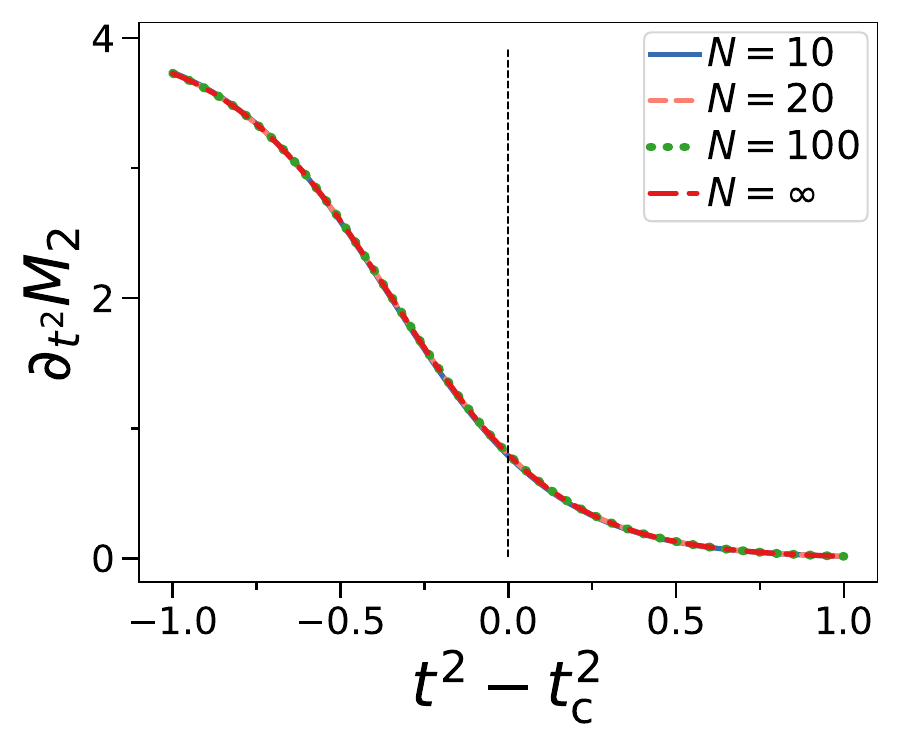}
	\caption{Universal behavior of $2$-SRE for random basis evolution model~\eqref{eq:random_CliffGUE}. \idg{a} $M_2$ against time $t$ for different $N$.  \idg{b} Derivative in respect to $t^2$ of SRE  $\partial_{t^2} M_2$ against $t$. Black vertical dashed line is critical time  $t_\text{c}^2=\frac{1}{4}N\ln(2)$.
 \idg{c} $\partial_{t^2} M_2$ against $t^2$ shifted by $t_\text{c}^2$. 
 Hallmark of universality is the crossing of all curves at the critical point $t_\text{c}^2$ and collapse to a single curve.
	}
	\label{fig:randomEvol_renyi2}
\end{figure*}

For $t=0$, Eq.~\ref{eq:random_CliffGUE} gives a Clifford state $\ket{0}$, while for $t\sim \sqrt{N}$ the SRE converges to the average value of the SRE for Haar random states.  
We compute $M_2$ for~\eqref{eq:random_CliffGUE} analytically using the result of Ref.~\cite{leone2022stabilizer}
\begin{align*}
&M_2(\theta,d,N)=-\ln[(3+2^N)^{-1}(4+(2^N-1)\times\label{eq:renyiGUE2}\numberthis\\
&\left[\frac{7\cdot 2^{2N}-3\cdot2^N+2^N(2^N+3)\cos(4\theta)-8}{8(2^{2N}-1)}\right]^d)]\,.
\end{align*}
In the limit of $N\gg1$, $d\gg N$ and using $\theta=\frac{2t}{\sqrt{d}}$, we find approximately
\begin{equation}\label{eq:Renyi2Approx}
M_2(t)\approx N\ln(2)-\ln(4+2^Ne^{-4t^2})
\end{equation}
giving us 
\begin{equation}\label{eq:M2Pauliapprox}
    M_2(t)\approx 4t^2
\end{equation}
for $t\ll \sqrt{N}$.
We confirm this scaling in Fig.\ref{fig:randomEvol_renyi2}a.

Similar to circuits with Clifford and T-gates, we find a transition in the SRE when it converges to its maximum. 
In particular, we observe that the derivative $\partial_{t^2}M_2(t)$ in respect to $t^2$   intersects at $t_c^2=\frac{1}{4}N\ln(2)$ for all $N$ as shown in Fig.\ref{fig:randomEvol_renyi2}b. We observe that the curves collapse onto a single line when shifted by $t_\text{c}^2$ in Fig.\ref{fig:randomEvol_renyi2}c, demonstrating its scale-invariant behavior.

Finally, we study the circuit model for the min-relative entropy of magic~\eqref{eq:dmin}. 
From numerically studies of our deep circuit model up to $N\le5$, we find that 
\begin{equation}\label{eq:dminapprox}
D_\text{min}(t)\approx t^2
\end{equation}
up to a time $t_\text{c}^2\approx N\ln(2)$, where it then converges to the average value of Haar random states (see Appendix~\ref{sec:dmin}).
From the bound~\eqref{eq:boundDmin} we have for $\alpha=2$~\cite{haug2023stabilizer}
\begin{equation}\label{eq:stabbound}  
D_\text{min}(t)\ge \frac{1}{4}M_2(t)\approx t^2\,,
\end{equation}
where as last step we inserted~\eqref{eq:M2Pauliapprox} for the $N\gg1$ limit. Comparing~\eqref{eq:dminapprox} and~\eqref{eq:stabbound}, we find that the bound is approximately saturated, demonstrating that~\eqref{eq:stabbound} is indeed a tight bound and cannot be improved further.

\section{Random Hamiltonian evolution}\label{sec:GUEevolve}
Next, we study the evolution of states under random Hamiltonians~\cite{cotler2017chaos}.
We evolve an initial random stabilizer state $\ket{\psi(0)}=\ket{\psi_\text{STAB}}$ state 
\begin{equation}\label{eq:GUEevolve}
    \ket{\psi(t)}=e^{-iHt}\ket{\psi(0)}
\end{equation} 
in time $t$. The Hamiltonian $H$ is chosen as a random matrix sampled from the Gaussian unitary ensemble (GUE).

We now calculate the SRE for the evolution with the random Hamiltonians.
First, we define the fidelity $F$ with the initial stabilizer state
\begin{equation}
F=\vert\braket{\psi(0)\vert\psi(t)}\vert^2
\end{equation}
For $t\ll1$, we find up to second order in $t$
\begin{align*}
F(t)\approx& \vert\bra{\psi}1-iHt-\frac{1}{2}H^2t^2\ket{\psi}\vert^2\approx\\
&1-t^2(\bra{\psi}H^2\ket{\psi}-\bra{\psi}H\ket{\psi}^2)
\end{align*}
We now normalize $H$ such that $\bra{\psi}H^2\ket{\psi}-\bra{\psi}H\ket{\psi}^2=1$ on average for $\ket{\psi}$ chosen from $2$-designs, which is achieved by demanding that on average one has $\text{tr}(H^2)=2^N+1$. 
This normalization factor can be computed exactly via the fact that $2$-designs have on average $\bra{\psi}H^2\ket{\psi}=\frac{\text{tr}(H^2)}{2^N}$ and $\bra{\psi}H\ket{\psi}^2=\frac{\text{tr}(H^2)}{2^N(2^N+1)}$. 
This restricts the eigenvalue spectrum of $H$ within $[-2,2]$ independent of $N$. This leads to am $N$-independent growth of correlations as proposed in Ref.~\cite{cotler2017chaos}. 

With this normalization of $H$, we get on average for short times $t\ll1$
\begin{equation}
F(t)\approx1-t^2\,.
\end{equation}
Due to Levy's lemma, observed expectation values such as $F(t)$  concentrate with exponentially high probability around its average for each sampled state~\cite{popescu2005foundations}.

\emph{Approximation of Pauli distribution.}
We now want to find an approximation for $M_\alpha(t)$ as function of time $t$. 
For this, we need to understand the distribution of expectation values $\beta_\sigma(t)\equiv\beta_\sigma=\bra{\psi(t)}\sigma\ket{\psi(t)}$ of $\ket{\psi(t)}$.
The distribution $\beta_\sigma^2=\bra{\psi}\sigma\ket{\psi}^2$ is the Pauli spectrum, i.e. the distribution of the square of Pauli string expectation values. In total, there are $4^N$ Pauli strings $\sigma$.
Any state can be written as $\rho=2^{-N}\sum_{\sigma}\beta_\sigma \sigma$, where $2^{-N}\sum_\sigma\beta_\sigma^2=1$ for pure states.

We have an initial stabilizer state $\ket{\psi(0)}$, which is stabilized by a commuting subgroup $G$ of $\vert G\vert=2^N$ Pauli strings. For any $\sigma\in G$, we have $\beta_\sigma^2=1$. In contrast, for $\sigma'\notin G$ we have $\beta_{\sigma'}^2=0$, where the complement of $G$ contains $4^N-2^N$ Pauli strings. 

\begin{figure*}[htbp]
	\centering	
	\subfigimg[width=0.26\textwidth]{a}{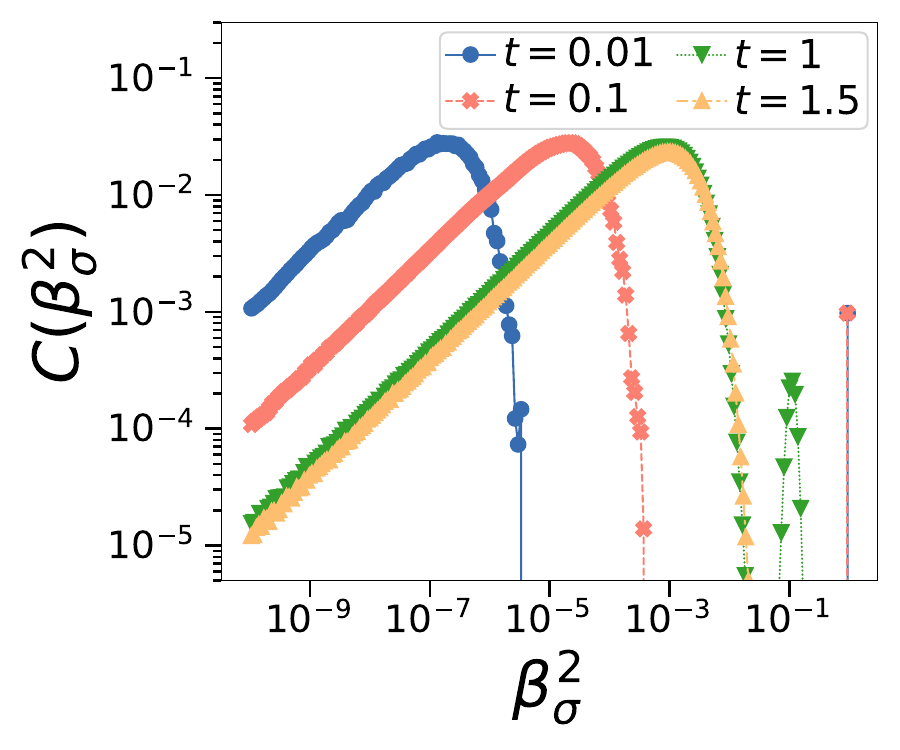}
\subfigimg[width=0.26\textwidth]{b}{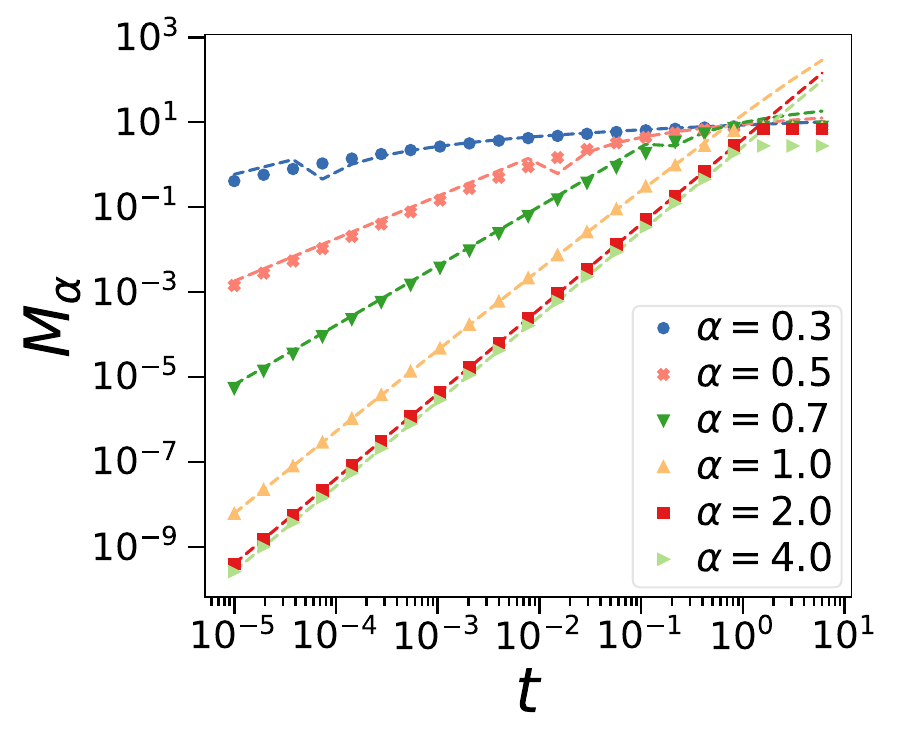}
\subfigimg[width=0.26\textwidth]{c}{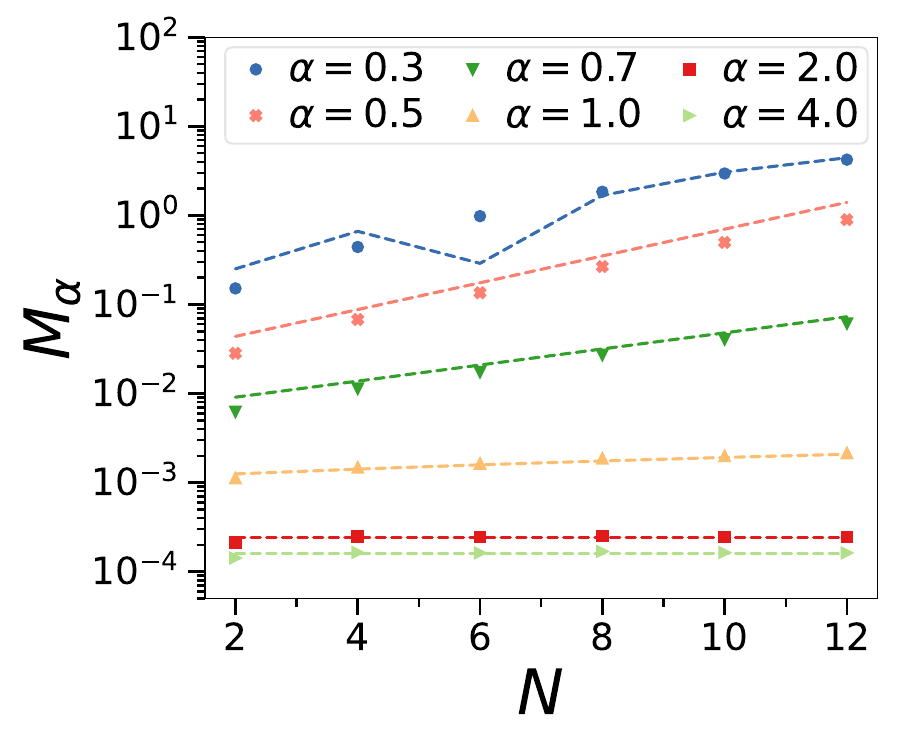}
	\caption{ $\alpha$-SREs of random Hamiltonian evolution.  \idg{a} Pauli spectrum plotted as histogram, where we show the probability $C$ of observing the Pauli expectation values $\beta_\sigma^2=\bra{\psi}\sigma\ket{\psi}^2$. We show different $t$ of GUE evolution for $N=10$ averaged over 20 random instances. For clarity, we do not show the trivial identity operator $\bra{\psi}I\ket{\psi}^2=1$ of the spectrum. \revA{For $t\gg1$, Pauli spectrum is mostly concentrated around $\beta_\sigma^2=2^{-N}\approx 10^{-3}$. }
 \idg{b} SRE $M_\alpha$ \revA{against $t$ for different $\alpha$ and $N=12$}, where dashed lines are approximations~\eqref{eq:RenyiAlphaGtr1},~\eqref{eq:RenyiAlphaLess1tsmall},~\eqref{eq:RenyiAlphaLess1tlarge} and~\eqref{eq:RenyiAlpha1}. The sudden change of the dashed line for $\alpha<1$ is due to the change from~\eqref{eq:RenyiAlphaLess1tsmall} to~\eqref{eq:RenyiAlphaLess1tlarge}  at $2^{N(1-\alpha)}(1-F^2)^\alpha=1$.
 \idg{c} SRE $M_\alpha$ against $N$ for $t=10^{-2}$. Dashed line is fit with approximations.
 Our model accurately describes $\alpha$-SREs of random evolution as function of $N$ and $t$, allowing us to predict overall scaling and critical time $t_{\text{c},\alpha}$. 
	}
	\label{fig:histPauli}
\end{figure*}

Now, how does the Pauli spectrum $\beta_\sigma(t)^2$ change when the stabilizer state is evolved in time $t$? 
For $t=0$, the Pauli spectrum has two peaks at $\beta_\sigma^2=0$ and $\beta_\sigma^2=1$. For $t>0$, the two peaks shift and diffuse. However, we observe numerically that the two peaks remain highly concentrated even for relatively large $t$. Note that for Haar random states, the concentration has been proven~\cite{turkeshi2023pauli}. 
We show the histogram of the Pauli spectrum in Fig.~\ref{fig:histPauli}a Note that up to $t\lesssim 1$, there are two distinct peaks with a gap in between them. Note that Fig.~\ref{fig:histPauli}a is a logarithmic plot, and the peak for small $\beta_\sigma^2$ appears broad in logarithmic space, but is actually very concentrated close to its mean value.
Let us now approximate the two peaks as Delta-functions centered around their mean value. For many qubits $N\gg1$, we can easily compute the average of each peak, i.e.
$\beta_{\sigma\in G}^2\approx F^2$, and $\beta_{\sigma\notin G}^2\approx 2^{-N}(1-F^2)$.

With decreasing $F$, the gap between $\beta_{\sigma\notin G}^2$ and $\beta_{\sigma\in G}^2$ decreases, and the two distributions merge when $F(t)^2\lesssim 2^{-N}$. As we will find, this happens at the critical time.
We confirm numerically that different instances of $H$ sampled from the GUE show similar spectrum.

\emph{SRE of random evolution.}
We now approximate the Pauli spectrum of $\ket{\psi(t)}$ by its two observed mean values. First, we split $M_\alpha$ into its contribution stemming from Pauli strings in $\sigma\in G$ and $\sigma \notin G$.
\begin{align*}
M_\alpha=&(1-\alpha)^{-1}\ln(2^{-N} \sum_\sigma \vert\beta_\sigma\vert^{2\alpha})=\\
&(1-\alpha)^{-1}\ln(2^{-N}[\sum_{\sigma\in G} \vert\beta_\sigma\vert^{2\alpha}+\sum_{\sigma\notin G} \vert\beta_\sigma\vert^{2\alpha}])
\end{align*}
Next, we approximate $\beta_{\sigma\in G}^2= F^2$ and $\beta_{\sigma\notin G}^2= 2^{-N}(1-F^2)$ and use that $\vert G\vert=2^N$ and $\vert \bar{G}\vert=4^N-2^N\approx 4^N$ for $N\gg1$, yielding our main approximation for the SRE
\begin{equation}\label{eq:main_fid_magic}
M_\alpha(F)\approx(1-\alpha)^{-1}\ln(F^{2\alpha}+2^{N(1-\alpha)}(1-F^2)^\alpha)
\end{equation}
Now, we regard the limit of $t\ll1$ and $N\gg1$. Here, we apply the first order Taylor expansions $F(t)\approx 1-t^2$, $1-F(t)^2\approx 2t^2$ and $\ln(F(t))\approx -t^2$ and insert them into~\eqref{eq:main_fid_magic}.

First, we study $\alpha<1$. We first demand that $2^{N(1-\alpha)}(1-F^2)^\alpha\ll1$ or $t\ll \frac{1}{\sqrt{2}}2^{-N(1-\alpha)/(2\alpha)}$, i.e. exponentially small times
\begin{equation}
M_{\alpha<1}\approx(1-\alpha)^{-1}2^{N(1-\alpha)}(1-F^2)^\alpha\approx \frac{2^\alpha}{1-\alpha} 2^{N(1-\alpha)}t^{2\alpha}\label{eq:RenyiAlphaLess1tsmall}\,.
\end{equation}
The growth in $M_{\alpha<1}$ is polynomial in $t$ and exponential in $N$.
Beyond exponentially small times $2^{N(1-\alpha)}(1-F^2)^\alpha\gg1$ and $t^2\le \frac{1}{2}$, we get for $\alpha<1$
\begin{align*}
M_{\alpha<1}\approx&(1-\alpha)^{-1}(N(1-\alpha)\ln(2)+\alpha\ln(1-F^2))\approx\\
&\frac{\alpha}{1-\alpha}\ln(2t^2)+N\ln(2)\,.\numberthis\label{eq:RenyiAlphaLess1tlarge}
\end{align*}
In particular, for $t\sim 1/\text{poly}(N)$, we find extensive $M_{\alpha<1}\sim N$.

Next, we regard the case $\alpha=1$, $t^2\le \frac{1}{2}$ and $N\gg1$. Here, we have
\begin{align*}
&M_{1}=2^{-N}\sum_\sigma \beta_\sigma^2\ln(\beta_\sigma^2)=\\
&-F^2\ln(F^2)-(1-F^2)\ln(2^{-N}(1-F^2))=\\
&2F^2\ln(F)+N(1-F^2)\ln(2)-(1-F^2)\ln(1-F^2)
\approx\\
&2(1-t^2)^2t^2+2Nt^2\ln(2)-2t^2\ln(2t^2)\approx\\
&2t^2(N\ln(2)-\ln(2t^2))\numberthis\label{eq:RenyiAlpha1}\,.
\end{align*}

Finally, we study the case $\alpha>1$, where we find
\begin{equation}\label{eq:RenyiAlphaGtr1}
M_{\alpha>1}\approx(1-\alpha)^{-1}\ln(F^{2\alpha})=\frac{2\alpha}{1-\alpha}\ln(F)\approx \frac{2\alpha}{\alpha-1}t^2\,.
\end{equation}
where we highlight that the growth is independent of $N$.
For $\alpha=2$ we have $M_2\approx 4t^2$, matching the result for the evolution in random bases of~\eqref{eq:M2Pauliapprox}. Also note that by comparing with~\eqref{eq:boundDmin} it is easy to see that all $\alpha>1$ provide tight lower bounds to $D_\text{min}$.

Our analytic results match numerical simulations as shown in Fig.~\ref{fig:histPauli}b,c for all investigated $\alpha$ and $N$. While we assumed small $t$ for the approximations, we observe that our equations match our numerical studies until the critical time where the SRE becomes maximal.
We show additional numerical results and the error relative to the approximations in Appendix~\ref{sec:numerics}.

\emph{Critical time.}
We now estimate the critical time $t_{\text{c},\alpha}$ for evolution with random Hamiltonians using our approximation. While these equations were derived for the limit of small $t$, our numeric suggest that the approximations work well up to the critical time when the SRE becomes maximal.
We define the critical time $t_{\text{c},\alpha}$ as the time when the SRE converges to its maximal value, i.e. $M_\alpha(t_{\text{c},\alpha})=M_{\alpha}^\text{max}$, where $M_{\alpha}^\text{max}$ has been computed in~\eqref{eq:renyimax} and we consider $N\gg1$. 
We now study $t_{\text{c},\alpha}$ for different $\alpha$ and its scaling with $N$.

First, SRE for $\alpha=0$ as given by~\eqref{eq:M0} relates to the number of Pauli expectation values which are exactly zero. The GUE evolution evolves all elements of the Pauli spectrum non-trivially and makes them non-zero with overwhelming probability, thus we get $\bra{\psi(t)}\sigma\ket{\psi(t)}^2\ne0$ for $\sigma \in\mathcal{P}$ for any $t>0$. Thus, the critical time is at 
\begin{equation}
    t_{\text{c},0}= 0\,,
\end{equation} 
matching the divergence observed in~\eqref{eq:RenyiAlphaLess1tsmall}. 
Next, we study $0<\alpha<1$. Here, inserting~\eqref{eq:RenyiAlphaLess1tlarge} into $M_\alpha(t_{\text{c},\alpha})=M_{\alpha}^\text{max}$ gives us 
\begin{equation}
    t_{\text{c},\alpha}^2\approx\frac{1}{2}\,.
\end{equation}
Most importantly, we find that the critical time is independent of $N$.  

Finally, for $\alpha>1$ we find using~\eqref{eq:RenyiAlphaGtr1} that the critical time grows linearly in $N$ 
\begin{align*}
1<\alpha\le2: \,\,\,t_{\text{c},\alpha}^2\approx&-\frac{1-\alpha}{2\alpha}N\ln(2)\,,\\
    \alpha>2: \,\,\,t_{\text{c},\alpha}^2\approx&\frac{1}{2\alpha}N\ln(2)\,.\numberthis
\end{align*}

Note that there may be constant corrections to $t_{\text{c}}$ not captured by our first-order approximations. However, we argue that the scaling of $t_{\text{c}}$ with $N$ is accurately captured by our approximations, as we get a good match between our derived formulas and numerical studies.
Our approximations were derived with the first order approximation of the fidelity $F\sim 1-t^2$. We  numerically study the behavior of $F$ for larger $t$. We find $F\sim e^{-t^2}$ up to a time $t\sim \sqrt{N}$. When inserting $F\sim e^{-t^2}$ into~\eqref{eq:main_fid_magic}, we also get~\eqref{eq:RenyiAlphaGtr1}, indicating that~\eqref{eq:RenyiAlphaGtr1} is indeed valid up to  $t\sim \sqrt{N}$. 
We note that at $\alpha=1$ a transition from constant to linear scaling occurs. We believe logarithmic corrections could appear here, however this warrants further studies.

Finally, as we show in Appendix~\ref{sec:GUEApp}, the Pauli spectrum and SRE of the GUE evolution matches closely the dynamics of the random basis evolution of Sec.~\ref{sec:Pauli}. We also observe that the SREs for both models match.
Thus, we argue that the scale-invariant behavior that we shown analytically in Sec.~\ref{sec:Pauli} for random bases evolution also emerges for the evolution with random Hamiltonians.

\section{Complexity and SRE}

We now show that SREs  can be related to the complexity of different operational tasks, where the type of task depends on $\alpha$.
In particular, we relate $\alpha>1$ to approximate fidelity estimation, while $\alpha<1$ to Clifford simulation complexity.

First, we note that for $\alpha>1$, the SRE is similar to $D_\text{min}$, which is the distance to the closest stabilizer state~\cite{haug2023efficient}. 
We find that the previously proven lower bound~\cite{haug2023stabilizer} is indeed tight for random evolutions~\eqref{eq:stabbound} which we numerically confirm in Appendix~\ref{sec:robustness}. Note that numeric evidence shows that SREs for $\alpha>1$ also provide an $N$-independent upper bound to $D_\text{min}$~\cite{haug2023efficient}. 
As such, we argue that SREs with $\alpha>1$ probe the closeness to the nearest stabilizer state. 

SREs with $\alpha=2$ have been shown an explicit operational meaning: They give a lower bound on fidelity estimation~\cite{flammia2011direct,leone2023nonstabilizerness}: Given a target state $\ket{\psi}$, one can estimate the fidelity with actual state $\rho$  by measuring Pauli expectation values~\cite{flammia2011direct}. The number of samples $m$ to estimate the fidelity is lower bounded as $m\gtrsim \exp(M_2(\ket{\psi}))$, while an upper bound is given by $m\lesssim \exp(M_0(\ket{\psi}))$~\cite{leone2023nonstabilizerness}.

In Appendix~\ref{sec:fidelity}, we show an explicit fidelity estimation algorithm whose efficiency is directly given by $M_2(\ket{\psi})$:  Fist, we note that fidelity estimation works especially well when $\ket{\psi}$ is a stabilizer state. Now, if $\ket{\psi}$ is not a stabilizer, but close to stabilizer state, then one can use this stabilizer state as a proxy for fidelity estimation: 
One estimates the fidelity of $\rho$ with the closest stabilizer state $\text{argmax}_{\ket{\psi_\text{STAB}}}\vert\braket{\psi\vert\psi_\text{STAB}}\vert^2$. We show that when $\ket{\psi}$ is close to a stabilizer state, this is a good approximation of the fidelity, where the quality depends on $M_2(\ket{\psi})$. 
In particular, we show in Appendix~\ref{sec:fidelity} that this scheme has an error $\Delta F \leq 2\sqrt{1-\exp(-M_2(\ket{\psi}))}$ and
gives a non-trivial approximation of the fidelity as long as $M_2(\ket{\psi})\leq \log(4/3)$.

In contrast, SREs with $\alpha<1$ (especially $\alpha=1/2$) show behavior similar to the log-free robustness of magic $\text{LR}$~\cite{howard2017application,rall2019simulation,liu2022many} or max-relative entropy of magic~\cite{bravyi2019simulation,liu2022many}. They respectively relate to the negativity of the mixture of stabilizer states, or the number of superpositions of stabilizer states needed to represent a given state. 
$\text{LR}$ has been used to estimate fault-tolerant state preparation complexity and relates to the complexity of Clifford based simulation algorithms. These algorithms simulate quantum circuits as Clifford circuits injected with nonstabilizer gates, where the simulation complexity commonly increase exponentially with the number of nonstabilizer gates~\cite{bravyi2016trading}. In fact, $M_{1/2}$ has been used as a proxy for $\text{LR}$ to evaluate simulation complexity of Clifford-based simulation algorithms~\cite{rall2019simulation}. Further, we find that the lower bound $\text{LR}\ge \frac{1}{2}M_{1/2}$~\eqref{eq:mn_lr} is tight for random evolution (see Appendix~\ref{sec:robustness}). 
Additionally, $M_0$ is a lower bound to the stabilizer nullity $\nu \geq M_0$, which characterizes the complexity of Clifford-based learning algorithms~\cite{gu2024magic, grewal2023efficient,hangleiter2023bell,leone2023learning}.

\emph{SRE dynamics and complexity.}
Now, we study the complexity of different tasks for doped Clifford circuits and random evolution using SREs.

For Clifford gates injected with T-gates, SREs converge for all $\alpha$ to their maximum $M_\alpha\sim N$ at T-gate density $q\sim \text{const}$. This is because each T-gate affects only a discrete subset of the Pauli spectrum.
We numerically find that $D_\text{min}$ and $\text{LR}$ appear to show this behavior as well. This implies that fidelity with the closest stabilizer state and classical cost of simulation with Clifford+T correlate. In particular, for $q\sim 1/N$ and thus $M_\alpha=\text{const}$, one can efficiently simulate and learn the state~\cite{grewal2023efficient,hangleiter2023bell,leone2023learning}, as well as estimate the fidelity~\cite{leone2023nonstabilizerness}.
In contrast, for $q\sim \text{const}$ and thus $M_\alpha\sim N$ simulation, learning and fidelity estimation is unlikely to be efficient~\cite{bravyi2016improved,bravyi2016trading,leone2021quantum}. 

In contrast, SREs for random evolution shows widely different behavior depending on $\alpha$. This is because random evolution affects all Pauli strings even at short evolution times. 
For $\alpha>1$, $M_{\alpha>1}$ grows as $\propto t^2$ and converges to its maximum at $t_{\text{c},\alpha>1}\sim \sqrt{N}$. For $t=\text{const}$, we have $M_{\alpha>1}=\text{const}$ and $D_\text{min}=\text{const}$, i.e. the evolution is close to a stabilizer state in terms of fidelity. Thus, at $t=\text{const}$ one can efficiently certify the fidelity of random evolution (see Appendix~\ref{sec:fidelity}).

$\alpha<1$ shows a completely different behavior, growing extremely fast with $t$:
For $\alpha=0$, $M_0$ is maximal for any $t>0$, which implies that stabilizer nullity $\nu$ is maximal, rendering known near-Clifford learning algorithms inefficient~\cite{grewal2023efficient,hangleiter2023bell,leone2023learning}.
For $0<\alpha<1$, SREs saturate rapidly at constant evolution time $t_{\text{c},0<\alpha<1}\approx \frac{1}{2}$. 
Further, extensive $M_{\alpha<1}\sim N$ is reached already for $t\gtrsim 1/\text{poly}(N)$.
This hints that simulating random dynamics to arbitrary precision with Clifford simulation algorithms becomes classically intractable already at $t\gtrsim 1/\text{poly}(N)$.

\begin{figure}[htbp]
	\centering	
	\subfigimg[width=0.3\textwidth]{}{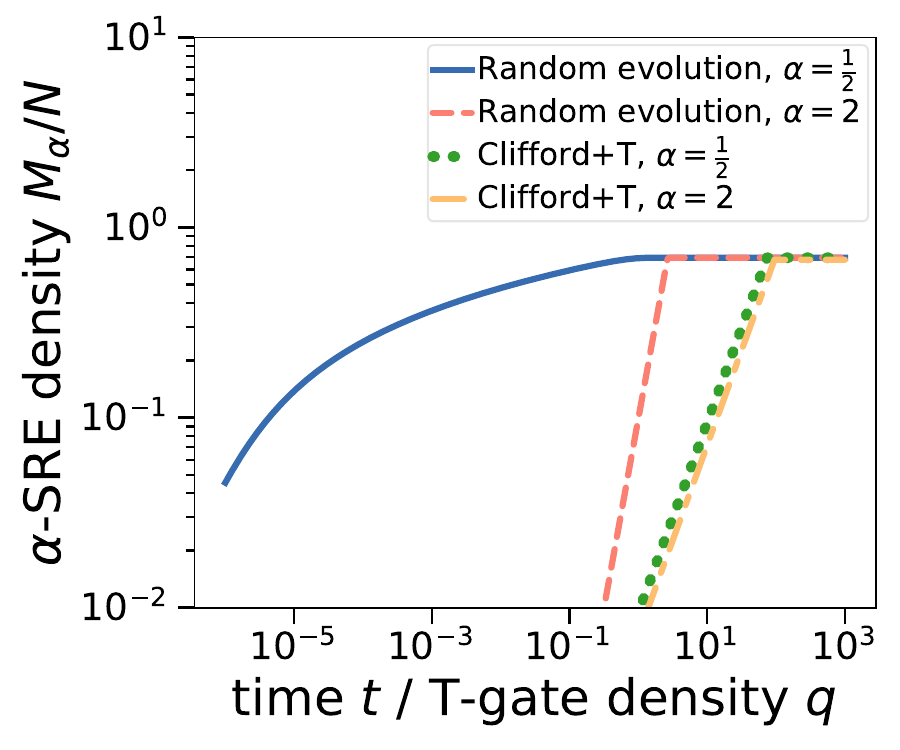}
	\caption{SRE of random Hamiltonian evolution for $\alpha<1$ increases extremely fast with time $t$. In contrast, for $\alpha>1$, as well as Clifford + T circuits for any $\alpha$, the growth in SRE is much slower. 
    We show SRE density $M_{\alpha}/N$ for random Hamiltonian evolution with~\eqref{eq:main_fid_magic} and doped Clifford circuits with~\eqref{eq:SRECliffordNT} for $N=40$.
	}
	\label{fig:SRE_Evolve_CliffordT_Alpha}
\end{figure}

We summarize the growth of SRE for random Hamiltonian evolution with time $t$, and circuits composed of Clifford unitaries and T-gates with T-gate density $q$ in Fig.~\ref{fig:SRE_Evolve_CliffordT_Alpha}. In particular, we highlight the extremely fast growth in time $t$ for random evolution for $\alpha<1$, which is not observed for $\alpha>1$ or for Clifford+T circuits for any $\alpha$.

Finally, we note that the complexity of tasks related to entanglement tasks can be bounded using the stabilizer nullity~\cite{gu2024magic}. For small stabilizer nullity, one can efficiently do entanglement estimation, distillation and dilution. The same tasks become hard when the stabilizer nullity becomes extensive. 
For Cliffords doped with a constant number of T-gates we have a constant stabilizer nullity, and thus entanglement-related tasks are easy. In contrast, for random evolution the stabilizer nullity becomes already maximal for exponentially small times, where entanglement-related tasks become hard.

\section{Discussion}
We have studied $\alpha$-SREs for random Clifford circuits doped with T-gates and random time evolution where we demonstrated the connection of R\'enyi index $\alpha$ to different aspects of complexity of quantum states. 

We find that the SRE converges to the maximum at a critical T-gate density $q_{\text{c},\alpha}$ and time $t_{\text{c},\alpha}$ in the thermodynamic limit. We determine the transition exactly for $\alpha=2$, while for general $\alpha$ we determine the convergence using heuristic models of the Pauli spectrum.
For $\alpha=2$, we observe universal behavior around the critical point where the derivative of the SRE can be rescaled onto a single curve for all $N$.
This hints that the saturation transition is connected to phase transitions, where universal behavior is commonly found, for example for the transition between different phases of quantum many-body system~\cite{osterloh2002scaling} or at complexity transitions in classical and quantum algorithms~\cite{zhang2022quantum}.
\revA{While we can show the universality analytically for $\alpha=2$, a follow-up work observes strong indications of universality numerically for $\alpha=1$ by studying system sizes up to $N=24$~\cite{tarabunga2025efficientmutualmagicmagic}. We believe universality holds for all $\alpha$, which would be interesting to study in future works. For example, using recent methods for Clifford circuits doped with magic, it may be possible to find analytic expressions beyond $\alpha=2$~\cite{leone2025non}. }

The critical T-gate density $q_{\text{c},\alpha}$ shows non-monotonous behavior as function of R\'enyi index $\alpha$, and the critical evolution time $t_{\text{c},\alpha}$ even changes its scaling with qubit number $N$. This behavior highlights the fact that SREs with different $\alpha$, i.e. different moments of the Pauli spectrum $\bra{\psi}\sigma\ket{\psi}^{2\alpha}$, probe different aspects of nonstabilizerness.

SREs with $\alpha<1$ and $\alpha>1$ probe two different aspects of nonstabilizernes:
We find that SREs with $\alpha<1$ behave similar to the log-free robustness of magic $\text{LR}$~\cite{howard2017application,rall2019simulation,liu2022many} or max-relative entropy of magic~\cite{bravyi2019simulation,liu2022many}. Roughly, these magic monotones relate to the complexity of approximating a state using a superposition of stabilizer states. 
In contrast, $\alpha>1$ SREs can be related~\cite{haug2023efficient} to the min-relative entropy of magic $D_\text{min}$~\cite{bravyi2019simulation,liu2022many}, which measure the fidelity with the closest stabilizer state.

We find that random Hamiltonian evolution has a fundamental separation in nonstabilizerness complexity: Simulating random evolution using Clifford-based algorithms to arbitrary precision is already hard for very small times $t\gtrsim 1/\text{poly}(N)$. In contrast, up to $t=\text{const}$ one can efficiently certify the fidelity of the evolved state within some fixed error $\Delta F\approx 4t$. We show the algorithm in Appendix~\ref{sec:fidelity} by certifying in respect to a stabilizer approximation of the state. 
In contrast, for Cliffords doped with T-gates, simulation and certification complexity correlate, and become intractable for the same T-gate density $q\gtrsim\log(N)$.

Finally, for both random evolution and Clifford+T model, the critical time and T-gate density is maximal for $\alpha=2$. This indicates that the $2$-SRE holds a special status. Coincidentally, for $\alpha<2$, the SRE is known not to be a monotone~\cite{haug2023stabilizer}, while for $\alpha\ge2$ it is a pure state monotone~\cite{leone2024stabilizer}.

Finally, we want to highlight the technical contributions of our work which may be of independent interest: We show heuristically that the Pauli spectrum of random Hamiltonian evolution can be approximated by two distinct peaks. With increasing time, the two peaks shift towards each other and eventually merge. This is exactly when the SRE becomes maximal. 
At last, we introduce a class of random evolution in~\eqref{eq:random_CliffGUE}, which can be seen as evolution in random Clifford bases. This evolution  behaves very similar to random Hamiltonian evolution, where we observe numerically the same Pauli spectrum. It can be expressed as random Clifford circuits combined with small-angle single-qubit rotations. This allows us to compute its $2$-SRE analytically for all times $t$. The random Clifford bases evolution could serve as a model of random evolution with an exact circuit representation.

\begin{acknowledgements}
We thank Hyukjoon Kwon, Ludovico Lami and especially Lorenzo Piroli for inspiring discussions. We also thank Emanuel Dallas for noticing a typo in~\eqref{eq:main_fid_magic}.
This work is supported by a Samsung GRC project and the UK Hub in Quantum Computing and Simulation, part of the UK National Quantum Technologies Programme with funding from UKRI EPSRC grant EP/T001062/1. 
\end{acknowledgements}

\bibliography{bibliography}

\onecolumngrid
\newpage 

\appendix
\setcounter{secnumdepth}{2}
\setcounter{equation}{0}
\setcounter{figure}{0}
\renewcommand{\thetable}{S\arabic{table}}
\renewcommand{\theequation}{S\arabic{equation}}
\renewcommand{\thefigure}{S\arabic{figure}}

\clearpage
\begin{center}
	\textbf{\large Appendix}
\end{center}
\setcounter{equation}{0}
\setcounter{figure}{0}
\setcounter{table}{0}
\makeatletter
\renewcommand{\theequation}{S\arabic{equation}}
\renewcommand{\thefigure}{S\arabic{figure}}
\renewcommand{\bibnumfmt}[1]{[S#1]}
\newtheorem{thmS}{Theorem S\ignorespaces}

We provide additional technical details and data supporting the claims in the main text.

\revA{
\section{Comparison maximal magic and Haar random states}\label{sec:compHaar}
In the main text, we derive the maximal SRE $M_\alpha^\text{max}$ as function of $\alpha$. To derive this result, we approximate it with the SRE of a state that has effectively a completely uniform Pauli spectrum with $\bra{\psi}P\ket{\psi}=2^{-N}$ for $P\neq I$. 

We now provide an alternative way to compute the maximal magic.
It is known that Haar random states have in the limit of large $N$ the maximal possible magic~\cite{liu2022many}. The average SRE of Haar random states has been derived analytically in Ref.~\cite{turkeshi2023pauli} in the limit of large $N$ as
\begin{equation}
    M_\alpha^\text{Haar}=\frac{1}{1-\alpha}\ln\left[\frac{(4^N-1)(2(2^{N-1}+1)^{-1})^\alpha \Gamma(\alpha+\frac{1}{2})}{\sqrt{\pi}2^N}+2^{-N}\right]\,,
\end{equation}
where $\Gamma(x)$ is the Gamma function.
Now, we look at the asymptotics for large $N$, where we find
\begin{equation}
    M_\alpha^\text{Haar}\approx\frac{1}{1-\alpha}\ln\left[\frac{2^N2^{-N\alpha} \Gamma(\alpha+\frac{1}{2})}{\sqrt{\pi}}+2^{-N}\right]\,.
\end{equation}
Now, let us regard $\alpha>2$. Here, the second term dominates and we have
\begin{equation}
    M_{\alpha>2}^\text{Haar}\approx\frac{1}{1-\alpha}\ln\left[2^{-N}\right]=-\frac{1}{1-\alpha}N\ln2\equiv M_{\alpha>2}^\text{uniform}\,,
\end{equation}
matching the maximal magic we derived via the flat Pauli spectrum. 
Next, for $\alpha<2$, the first term dominates with
\begin{equation}
    M_{\alpha<2}^\text{Haar}\approx\frac{1}{1-\alpha}\ln\left[\frac{2^{N(1-\alpha)} \Gamma(\alpha+\frac{1}{2})}{\sqrt{\pi}}\right]= N\ln2 + \frac{1}{1-\alpha}\ln\left[\frac{\Gamma(\alpha+\frac{1}{2})}{\sqrt{\pi}}\right] \equiv M_{\alpha<2}^\text{uniform} + \text{const}\,,
\end{equation}
matching the flat Pauli spectrum result up to constant corrections.
}

\section{GUE evolution and random basis evolution}\label{sec:GUEApp}
We now give numeric evidence that the evolution via $\ket{\psi(t)}=\exp(-iHt)\ket{0}$ with a random Hamiltonian $H$ sampled from the GUE has on average the same Pauli spectrum as the evolution in random Clifford bases via $d\gg N$ single-qubit rotations with  parameters $\theta=2t/\sqrt{d}$ as defined in~\eqref{eq:random_CliffGUE}.

\begin{figure*}[htbp]
	\centering	
	\subfigimg[width=0.3\textwidth]{}{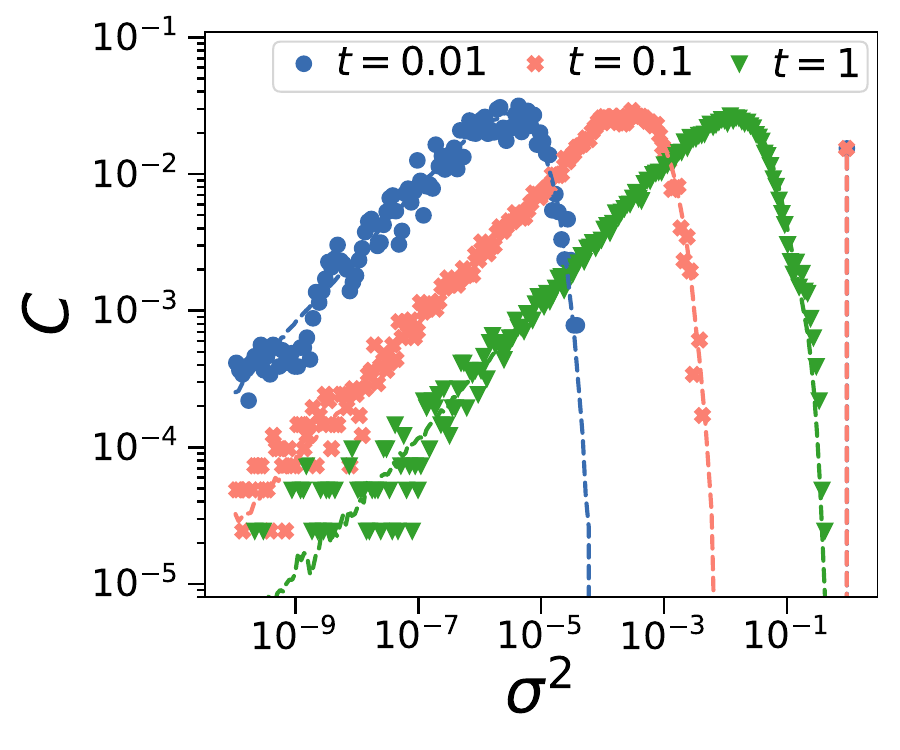}
	\caption{Pauli spectrum plotted as histogram over occurrences $C$ of observing Pauli expectation value $\langle\sigma\rangle^2$. The dots for evolution in random Clifford bases with rotations $\theta=2t/\sqrt{d}$ for different $t$. The dashed line is evolution with GUE Hamiltonian. We show $N=6$, $d=1000$ and average over 1000 random Hamiltonians and 10 circuits. For clarity, we do not show the trivial identity operator $\bra{\psi}I\ket{\psi}^2=1$ of the spectrum.
	}
	\label{fig:cliffordGUE}
\end{figure*}

In Fig.~\ref{fig:cliffordGUE}, we plot the Pauli spectrum, where $C(\bra{\psi}\sigma\ket{\psi}^2)$ is the probability  of finding Pauli expectation value $\bra{\psi}\sigma\ket{\psi}^2$ of Pauli $\sigma$ for a given state $\ket{\psi(t)}$. We show $C(\bra{\psi}\sigma\ket{\psi}^2)$ for different $t$ for evolution in random Clifford bases (dots) as well as the GUE evolution with same $t$ (dashed lines). We observe that both match nearly perfectly, indicating that they have the same statistical properties in terms of Pauli spectrum and SRE.

While we believe that both evolutions show similar behavior for polynomial times, we note that for very long times (on the scale of $t\sim 2^{N/2}$ ) both models likely show different behavior in terms of deep thermalization~\cite{cotler2017chaos}, as the GUE Hamiltonian evolution conserves energy while the other model does not. It has been noted that the ensemble of GUE evolutions forms an exact $k$-design at polynomial times, however stops being a $k$-design at exponential times. This behavior at long times is attributed to energy conservation of the evolution, which at long times leads to a dephasing due to the energy eigenvalues. For non-energy conserved dynamics this behavior at long times is not expected.
However, this difference at exponential times is evident in the $k$-design properties, however it may not be evident in the Pauli spectrum and SRE~\cite{haug2023efficient}. 

The study of this subtleties at exponential times is difficult numerically, and we leave a formal study of the statistical similarity between~\eqref{eq:random_CliffGUE} and evolution with random Hamiltonians as an open problem.

\section{Min-relative entropy of magic scaling for random evolution}\label{sec:dmin}
We show the min-relative entropy of magic $D_\text{min}$ as function of time $t$ for evolution with random Hamiltonians sampled from the GUE. We find that the increase with $t$ can be approximated by $D_\text{min}\approx t^2$ up to the time when it converges to its maximal value $D_\text{min}\le N\ln(2)$. 
\begin{figure}[htbp]
	\centering	
	\subfigimg[width=0.3\textwidth]{}{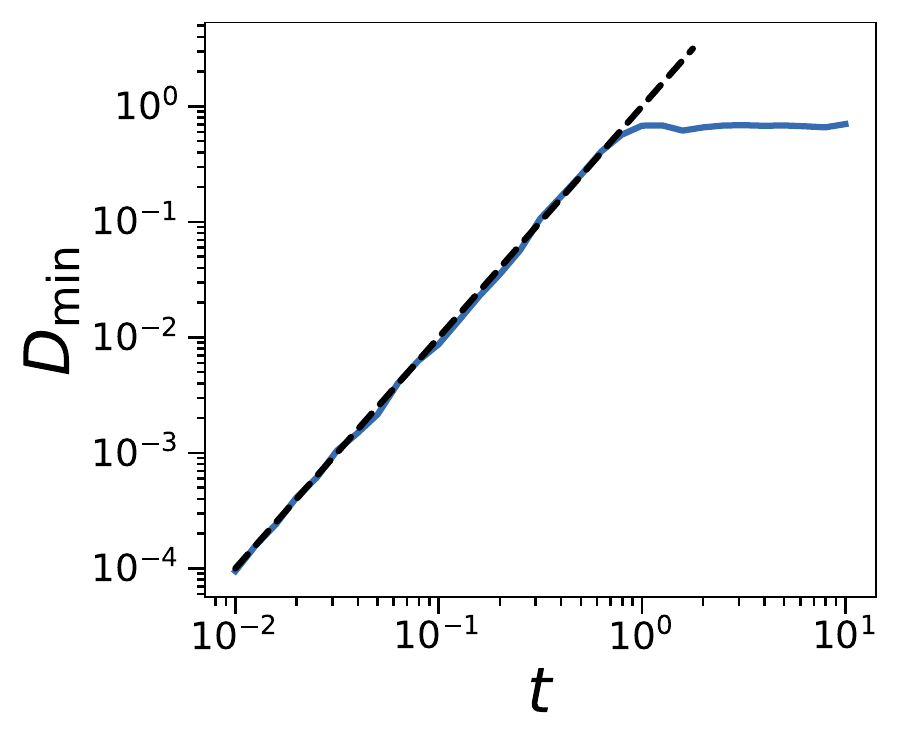}
	\caption{Min-relative entropy of magic $D_\text{min}$ as function of $t$ for evolution with GUE Hamiltonian. Dashed line is $t^2$. We show $N=4$.
	}
	\label{fig:randomDmin}
\end{figure}

\revA{
\section{Further numerical results}\label{sec:numerics}
Here, we show additional numerical results on the time evolution with random Hamiltonians.

First, in Fig.~\ref{fig:SREevolveT} we show SRE $M_\alpha(t)$ against time $t$ for different $\alpha$, where the dashed lines are the approximations $M_\alpha^\text{approx}$ derived in the main text. Here, we simulate the dynamics using  statevector simulation.
In Fig.~\ref{fig:SREevolveT}b, we show the error  $\vert M_\alpha - M_\alpha^\text{approx} \vert/M_\alpha$ relative to the approximations, where we find a close match for $\alpha>1$ for all $t$, while $\alpha>1$ matches only for $t<1$. Note that the approximations were derived for small $t$, yet we find that they work quite well even for larger $t$.
\begin{figure*}[htbp]
	\centering	
	\subfigimg[width=0.3\textwidth]{a}{mfitSEDmagicAN12.pdf}
\subfigimg[width=0.3\textwidth]{b}{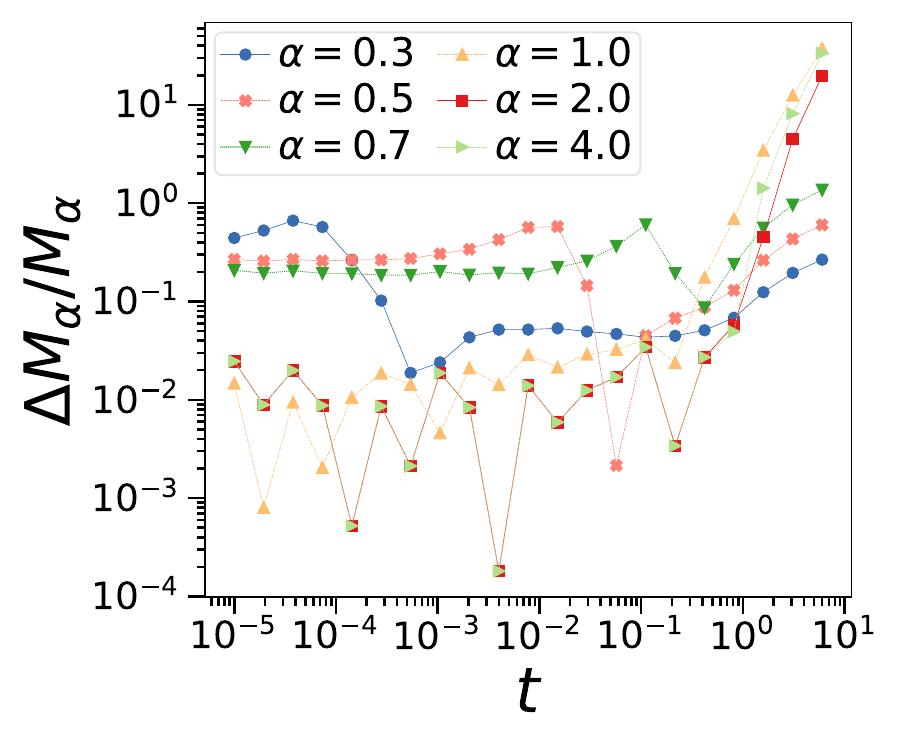}
	\caption{ $\alpha$-SREs of random Hamiltonian evolution. 
 \idg{a} SRE $M_\alpha(t)$ against time $t$ for $N=12$ and different $\alpha$, where dashed lines are approximations $m_\alpha^\text{approx}$ of~\eqref{eq:RenyiAlphaGtr1},~\eqref{eq:RenyiAlphaLess1tsmall},~\eqref{eq:RenyiAlphaLess1tlarge} and~\eqref{eq:RenyiAlpha1}. The sudden change of the dashed line for $\alpha<1$ is due to the change from~\eqref{eq:RenyiAlphaLess1tsmall} to~\eqref{eq:RenyiAlphaLess1tlarge}  at $2^{N(1-\alpha)}(1-F^2)^\alpha=1$.
 \idg{b} Relative error $\vert M_\alpha - M_\alpha^\text{approx} \vert/M_\alpha$ against $t$.
	}
	\label{fig:SREevolveT}
\end{figure*}

In Fig.~\ref{fig:SREevolveN}, we show the SRE $M_\alpha(t)$ against qubit number $N$ at different times $t$ under random Hamiltonian evolution.  We find that the error depends only weakly on $N$, which gives evidence that our approximations are accurate even for larger $N$.

\begin{figure*}[htbp]
	\centering	
	\subfigimg[width=0.3\textwidth]{a}{combFitScale0_01EDmagicA.pdf}
\subfigimg[width=0.3\textwidth]{b}{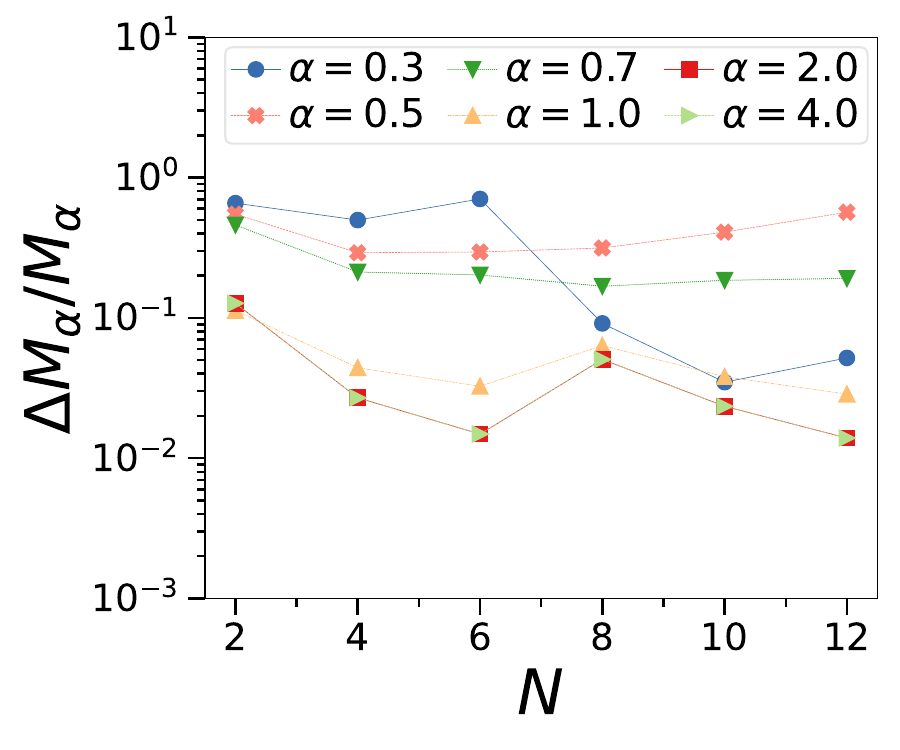}\\
	\subfigimg[width=0.3\textwidth]{c}{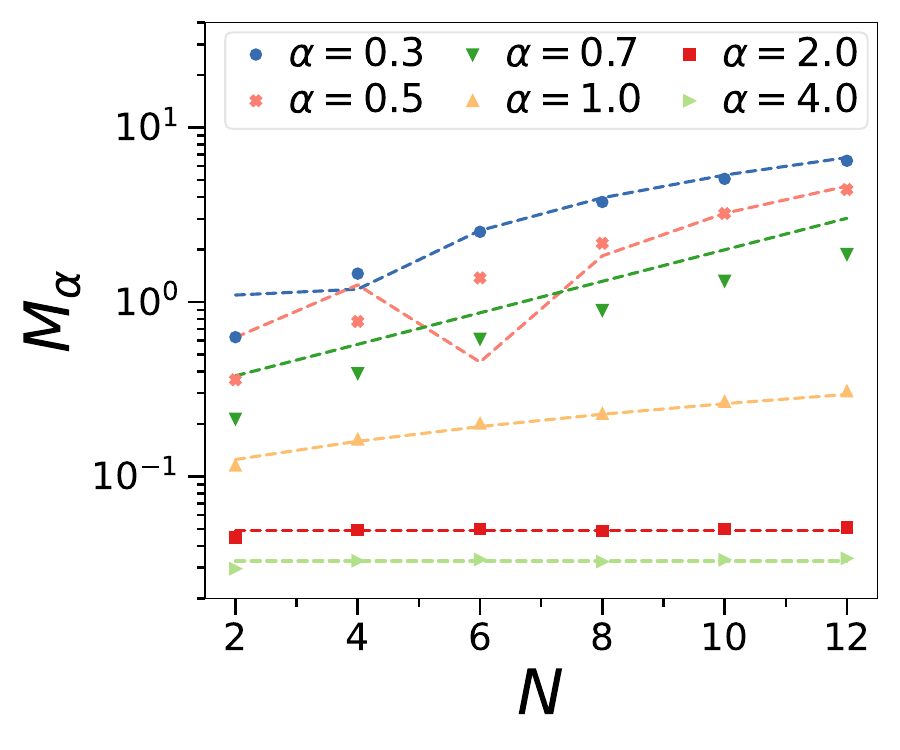}
\subfigimg[width=0.3\textwidth]{d}{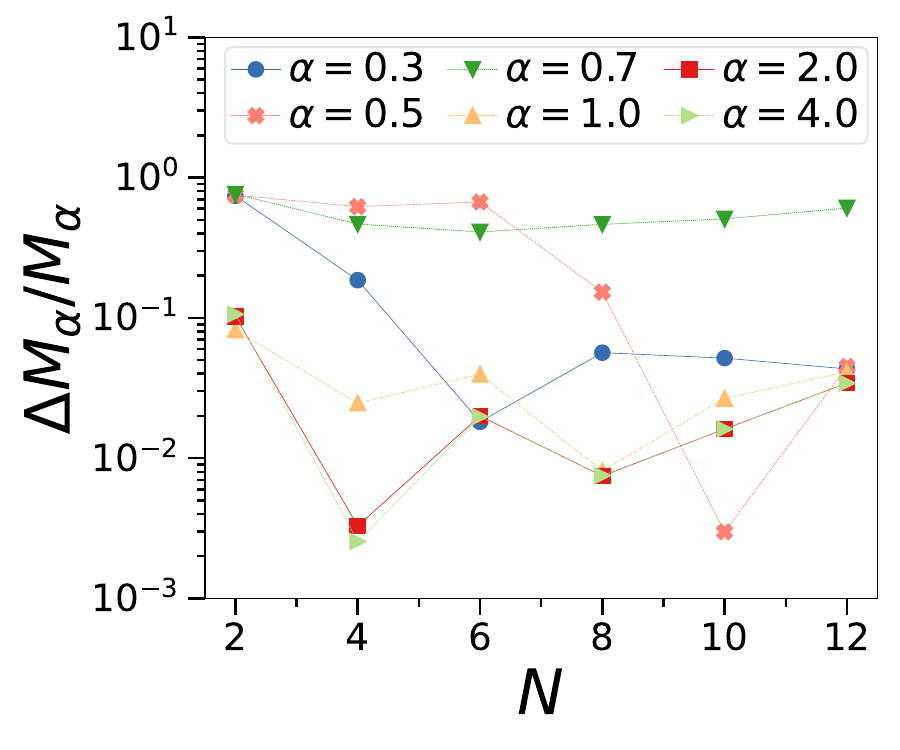}\\
	\subfigimg[width=0.3\textwidth]{e}{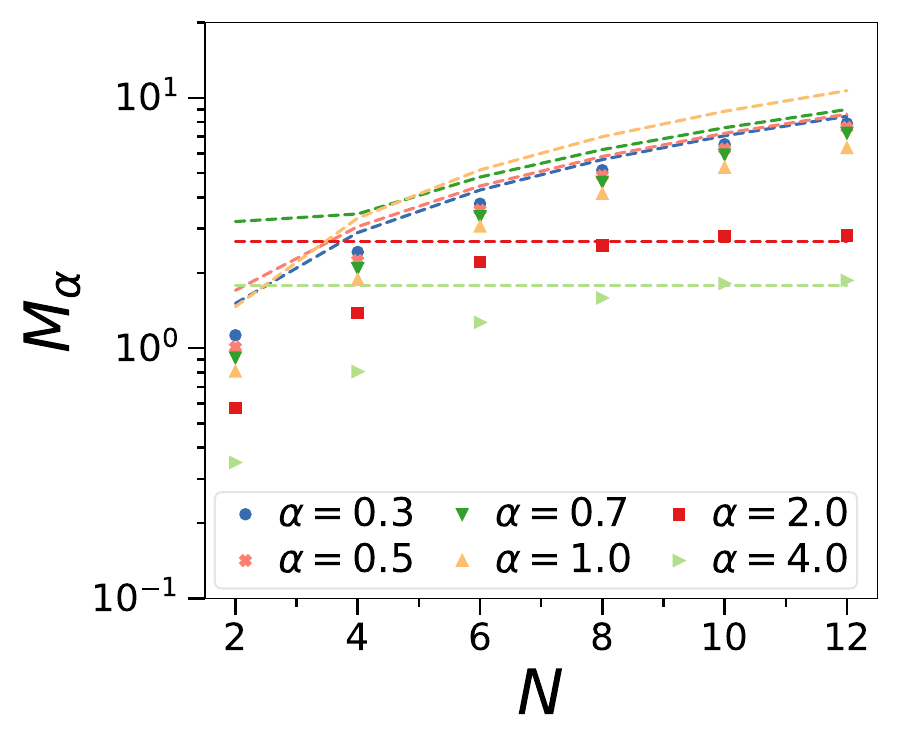}
\subfigimg[width=0.3\textwidth]{f}{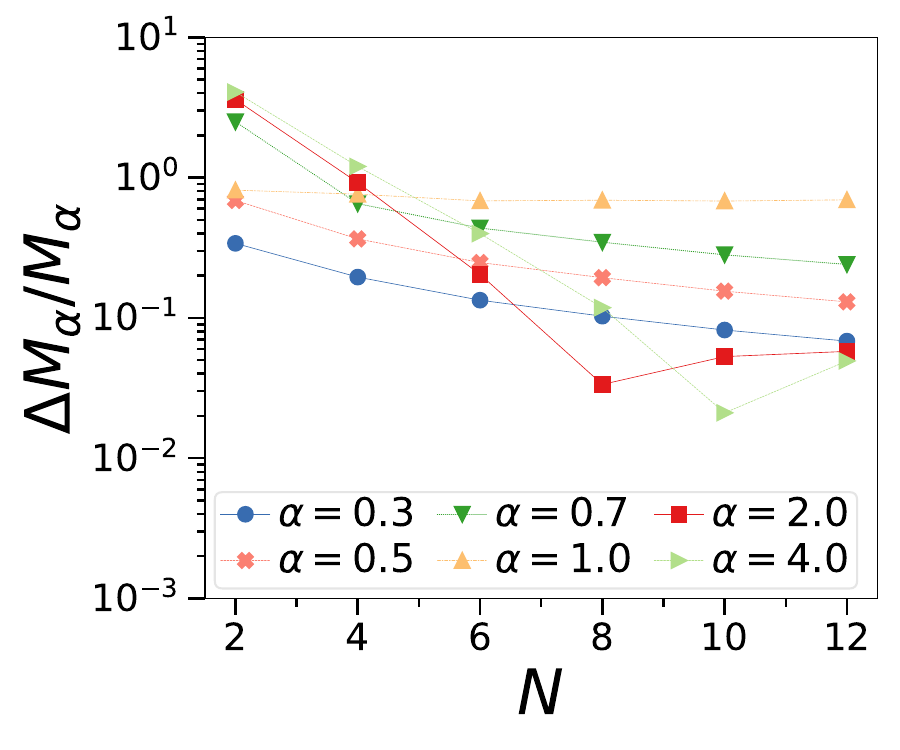}\\
	\caption{ $\alpha$-SREs of random Hamiltonian evolution against $N$ for different times \idg{a,b} $t=10^{-2}$, \idg{c,d} $t=10^{-1}$, and \idg{e,f} $t=1$.  \idg{a,c,e} SRE $M_\alpha(t)$ where dashed line is fit with approximations~\eqref{eq:RenyiAlphaGtr1},~\eqref{eq:RenyiAlphaLess1tsmall},~\eqref{eq:RenyiAlphaLess1tlarge} and~\eqref{eq:RenyiAlpha1}.  
    \idg{b,d,f} Relative error $\vert M_\alpha - M_\alpha^\text{approx} \vert/M_\alpha$ to approximations $M_\alpha^\text{approx}(t)$.
	}
	\label{fig:SREevolveN}
\end{figure*}

}

\section{SRE, min-relative entropy and robustness}\label{sec:robustness}
\begin{figure*}[htbp]
	\centering	
	\subfigimg[width=0.3\textwidth]{a}{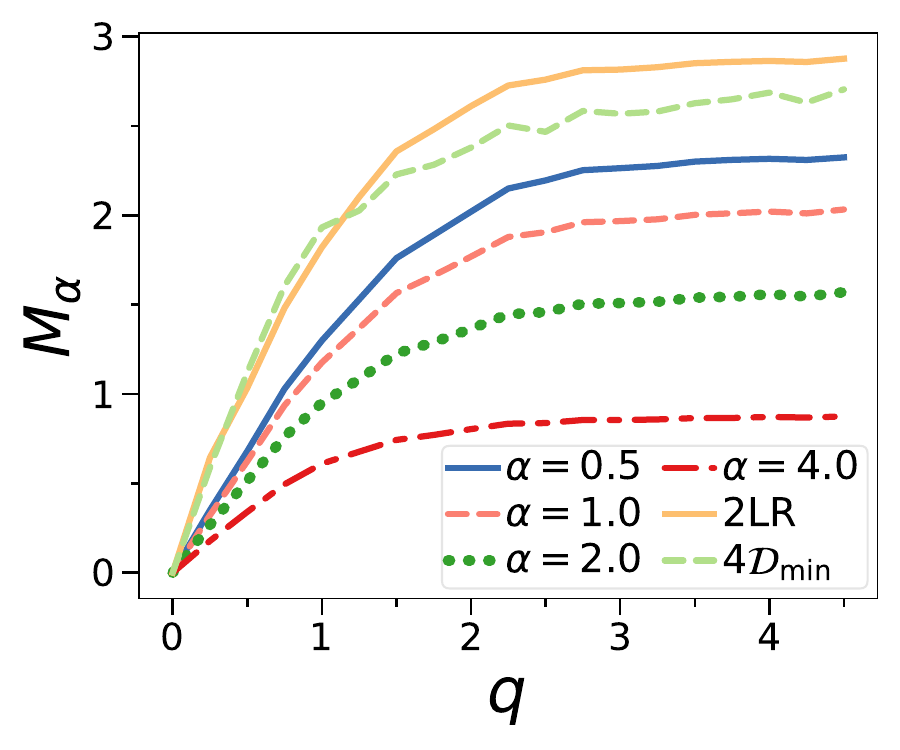}
 	\subfigimg[width=0.3\textwidth]{b}{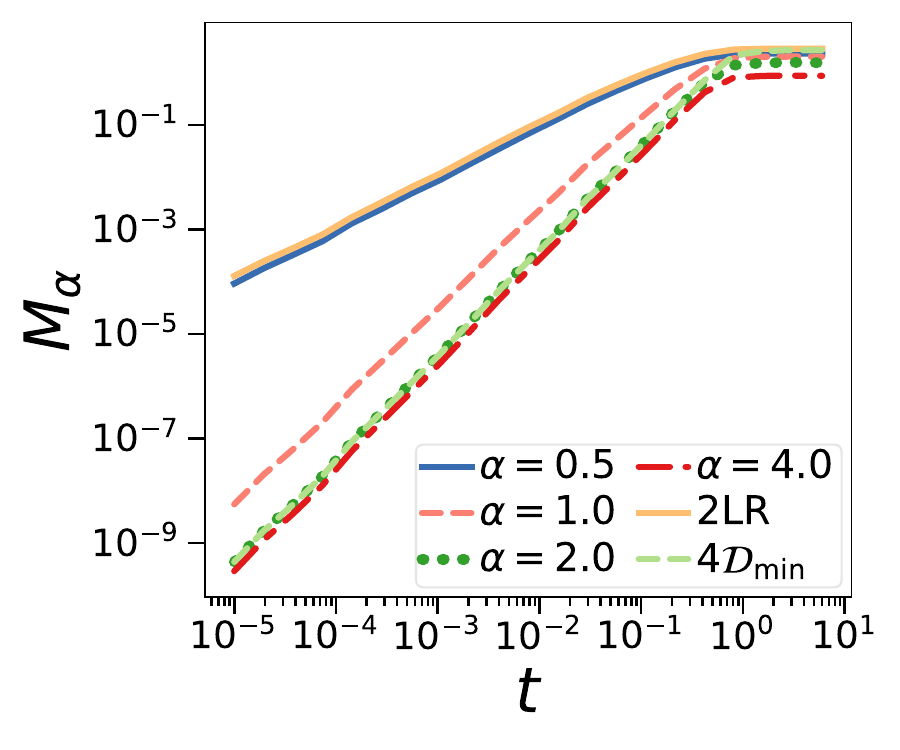}
	\caption{Relationship of $\alpha$-SRE with min-relative entropy $D_\text{min}$ and log-free robustness $\text{LR}$.  Here, we rescaled $D_\text{min}$ such that it poses an upper bound on SRE for $\alpha=2$, and $\text{LR}$ as upper bound for $\alpha\geq1/2$.
 \idg{a} Clifford+T circuit for $N=4$ qubits against T-gate density $q$.
  \idg{b} Evolution with random Hamiltonian sampled from GUE against time $t$.
	}
	\label{fig:robustness}
\end{figure*}

Here, we study the relationship between $\alpha$-SREs, min-relative entropy $D_\text{min}$ and log-free robustness $\text{LR}$.
We show in Fig.~\ref{fig:robustness}a the growth of $M_\alpha$, min-relative entropy $D_\text{min}$ and log-free robustness $\text{LR}$ with $N_\text{T}$. Here, we rescaled $D_\text{min}$ and $\text{LR}$ such that they correspond to their respective bounds, i.e. $2\text{LR}\geq M_{1/2}$ and $4D_\text{min}\geq M_{2}$. 
In Fig.~\ref{fig:robustness}a, we show the Clifford-T circuit, we find that $M_\alpha$ is indeed is a lower bound, which is non-tight.
In Fig.~\ref{fig:robustness}b, we show evolution with random Hamiltonian. Here, the lower bounds match closely, indicating that they are indeed tight.
We also note the relationship between $\text{LR}$, $D_\text{min}$ and $\alpha$. For $\alpha<1$, $\text{LR}$ and $M_\alpha$ show similar growth, indicating that they relate to classical simulation complexity. 
While for $\alpha>1$, $M_\alpha$ growth rate is similar to $D_\text{min}$ which measures the distance to the closest stabilizer state. 
We also note that the convergence to maximal $M_\alpha$ shows completely different scaling depending on $\alpha$, with $t_{\text{c},\alpha<1}^2=\text{const}$, while $t_{\text{c},\alpha>1}^2\propto N$. Note that this behavior is difficult to see for small $N$.

\section{State certification via Pauli measurements and SREs}\label{sec:fidelity}
A common task is state certification to check whether the prepared state $\rho$ is close to the ideal state $\ket{\psi}$ that one actually wanted to prepare. 
For this task, Ref.~\cite{flammia2011direct} proposed a simple algorithm that only requires to measure Pauli strings of the actual state. First, note that one can decompose any state in terms of its Pauli strings, i.e. $\rho=2^{-N}\sum_{\sigma\in\mathcal{P}}\beta_\sigma\sigma$ with Pauli expectation values $\beta_\sigma(\rho)=\text{tr}(\rho \sigma)$.
The fidelity between $\rho$ and $\ket{\psi}$ is given by
\begin{equation}
F(\rho,\ket{\psi})=\bra{\psi}\rho\ket{\psi}=2^{-N} \sum_{\sigma\in\mathcal{P}}\beta_\sigma(\ket{\psi})\beta_\sigma(\rho)
\end{equation}
Now, we note that $P_{\ket{\psi}}(\sigma)=2^{-N}\beta_\sigma(\ket{\psi})^2$ is a probability distribution for any pure state $\ket{\psi}$. We can rewrite the fidelity estimation into a sampling problem
\begin{equation}
F(\rho,\ket{\psi})=\sum_{\sigma\in\mathcal{P}}P(\sigma)\frac{\beta_\sigma(\rho)}{\beta_\sigma(\ket{\psi})}=\underset{\sigma \sim P_{\ket{\psi}}}{\mathbb{E}}[\frac{\beta_\sigma(\rho)}{\beta_\sigma(\ket{\psi})}]\,.
\end{equation}
Thus, to estimate $F$ we only need to sample from $P_{\ket{\psi}}(\sigma)$ and compute $\beta_\sigma(\ket{\psi})$ using some classical algorithm, and then measure the Pauli expectation value $\beta_\sigma(\rho)$ of the actual state $\rho$ on the quantum device. 

One can bound the number of Pauli measurements $m$ needed on the quantum computer using the SRE~\cite{leone2023nonstabilizerness}:
\begin{equation}
   \frac{2}{\epsilon^2}\ln(2/\delta)\exp(M_2(\ket{\psi})) \geq m \geq \frac{64}{\epsilon^4}\ln(2/\delta)\exp(M_0(\ket{\psi}))
\end{equation}
where $\epsilon$ is the additive accuracy and $\delta$ the probability the protocol fails. Most importantly, this algorithm has no assumptions on experimental state $\rho$, and only depends on properties of the reference state $\ket{\psi}$.
The protocol is always sample efficient when $M_0(\ket{\psi})=O(\log(N))$.
For example, stabilizer states can be certified with $O(1)$ samples.

Now, what happens for nonstabilizer states? From the lower bound, we know that the protocol becomes definitely inefficient when $M_2(\ket{\psi})=\omega(\log(N))$.
Thus, the protocol fails for the T-gate doped Clifford states for $q=\omega(\log(N))$.

The algorithm starts failing whenever one samples a $\sigma$ with  small, but non-zero magnitude $0<\vert\beta_\sigma(\ket{\psi})\vert<\gamma$ with some small threshold $\gamma$. 
From experiment, one estimates $\beta_\sigma(\rho)$ up to some additive error $\epsilon$. The resulting error is rescaled with the term in the denominator $\epsilon/\beta_\sigma(\ket{\psi})$.
Thus, to keep error low, one has to estimate  $\beta_\sigma(\rho)$ to high precision $\epsilon\sim\gamma$, which requires $m=1/\gamma^2$ samples. Thus, for $\gamma\sim 2^{-N}$ this results in an exponential cost.

Ref.~\cite{flammia2011direct} proposed an adapted protocol where one estimates the fidelity not in respect to $\ket{\psi}$, but in respect to a slightly perturbed state $\ket{\psi'}$ which does not feature Pauli expectation values with small, non-zero magnitudes.  
This incurs an error 
\begin{equation}
\vert F(\rho,\ket{\psi}) -F(\rho,\ket{\psi'}) \vert \leq \Vert \ket{\psi}\bra{\psi}- \ket{\psi'}\bra{\psi'}\Vert_2=\sqrt{2}\sqrt{1- \vert \braket{\psi\vert\psi'} \vert^2}\,.
\end{equation}
A good choice for the perturbed state $\ket{\psi'}$ is the closest stabilizer state to $\ket{\psi}$. In this case, we have
\begin{equation}
\vert F(\rho,\ket{\psi}) -F(\rho,\ket{\psi'}) \vert \leq \sqrt{2}\sqrt{1- F_\text{STAB}(\ket{\psi})}=\sqrt{2}\sqrt{1- \exp(-D_\text{min}(\ket{\psi})}\,.
\end{equation}
where $F_\text{STAB}$ is the stabilizer fidelity~\cite{bravyi2019simulation}.
$F_\text{STAB}$ is lower bounded by $M_2$ as shown in Ref.~\cite{haug2023efficient}
\begin{equation}
    F_\text{STAB}\geq 2\exp(-M_2)-1\,.
\end{equation}
Thus, we get
\begin{equation}
\Delta F=\vert F(\rho,\ket{\psi}) -F(\rho,\ket{\psi'}) \vert \leq 2\sqrt{1-\exp(-M_2(\ket{\psi}))}\,.
\end{equation}
There is a unique stabilizer state close to $\ket{\psi}$ as long as $F_\text{STAB}(\ket{\psi})>1/2$. Thus, the approximate certification can give a non-trivial result as long as $M_2\leq \ln(4/3)$. 

Now, let us  use the closest stabilizer state to certify the fidelity of random Hamiltonian evolution after time $t$. For evolved state $\ket{\psi(t)}$, for small $t$ the closest stabilizer state is $\ket{\psi(0)}$. 
We have $M_2(t)\approx 4t^2$ and thus certification error scaling as as $\Delta F\approx 4t$
Thus, one can certify the fidelity of $\ket{\psi(t)}$ with non-trivial error up to time $t\leq \frac{1}{4}$.

While $M_2$ is small for such $t$, note that this is not true for SREs with $\alpha<1$.  For example, $M_{1/2}(t)\approx \ln(2t^2)+N\ln(2)\sim N$ for any $t\gtrsim 1/\text{poly}(N)$. 

Commonly, Pauli fidelity certification becomes inefficient when states have a lot of nonstabilizerness~\cite{leone2023nonstabilizerness}. However, we argue that this statement applies strictly only for $\alpha>1$. This is because there are two different aspects of nonstabilizerness: While $\alpha<1$ relates to hardness of Clifford simulation, $\alpha>1$ measures the distance to the closest stabilizer states. 

For (approximate) Pauli fidelity estimation, where one checks the fidelity against the closest stabilizer state, the sampling complexity is related to the closest stabilizer state. As such, one can approximate the fidelity efficiently as long as $\alpha=2$ SRE $M_2$ is sufficiently small. This holds true even when the corresponding SRE with $\alpha<1$ SREs have already become extensive.

\end{document}